\documentclass[11pt,twoside]{article}  
\usepackage{mathptmx}
\usepackage{graphics,epsfig,latexsym,amssymb,amsbsy,amsmath}
\usepackage{multirow,booktabs}
\usepackage[commentmarkup=footnote]{changes} 
\newcommand{\blind}{1}

\usepackage[margin=10mm]{caption}

\usepackage{geometry}
\newcommand{\rP}{\mbox{P}}
\newcommand{\E}{\mbox{E}}
\newcommand{\rd}{\mbox{d}}
\newcommand{\var}{\mbox{var }}
\newcommand{\cov}{\mbox{cov }}

\newcommand*{\QED}{\hfill\ensuremath{\blacksquare}}%

\begin{document}

\if1\blind
{
\title{How mature are survival data at the time of an interim analysis \\ in a clinical trial with a survival outcome? }
 \author{}
  \author{Marianne A Jonker \thanks{corresponding author: marianne.jonker@radboudumc.nl}\hspace{.2cm}\\
    Department for Health Evidence, Section Biostatistics, \\
    Radboudumc, Nijmegen, The Netherlands\\
    and \\
    Steven Teerenstra \\
    Department for Health Evidence, Section Biostatistics, \\
   Radboudumc, Nijmegen, The Netherlands\\
     \\
    } 
}  \maketitle
 \fi

\if0\blind
{
  \bigskip
  \bigskip
  \bigskip
  \begin{center}
    {\LARGE\bf How mature are survival data at the time of an interim analysis \\ in a clinical trial with a survival outcome? }
\end{center}
  \medskip
} 
\fi

\maketitle

\abstract{
In a clinical trial with a survival outcome, an interim analysis is often performed to allow for early stopping for efficacy. If the interim analysis is early in the trial, one might conclude that a new treatment is more effective (compared to e.g.\ a placebo) and stop the trial, whereas the survival curves in the trial arms are not mature for the research question under investigation, for example because the curves are still close to 1 at that time. This means that the decision is based on a small percentage of the events in the long run only; possibly the events of the more frail patients in the trial who may not be representative for the whole group of patients. It may not be sensible to conclude effectiveness based on so little information. Criteria to determine the moment the interim analysis will be performed, should be chosen with care, and include the maturity of the data at the time of the interim analysis. Here, the expected survival rates at the interim analysis play a role. In this paper we will derive the asymptotic distribution of the Kaplan-Meier curves at the (random) moment the interim analysis will be performed for a one and two arm clinical trial. Based on this distribution, an interval in which the Kaplan Meier curves  will fall into (with probability 95\%) is derived and could be used to plan the moment of the interim analysis in the design stage of the trial, so before the trial starts.  }

\bigskip


\section{Introduction}
Suppose we plan a clinical trial with two arms, arm $A$ and arm $B$, and a survival (time-to-event) outcome. For example, cancer patients in arm $A$ get a new  treatment and those in arm $B$ a placebo or ``treatment as usual''. The effect of the new treatment is studied by comparing the overall survival or progression-free survival in the two arms. We consider trials where the aim is confirmatory testing (e.g., with the log-rank test). An example is a phase 3 trial since the focus is on confirming efficacy (and safety). The test for efficacy may require a substantial number of patients or substantial follow-up and therefore an interim analysis is often planned to allow for early stopping for efficacy. 
The moment the interim analysis is performed is determined based on one or more criteria which should be fully specified before the trial starts. For instance, the timing could be based on the number of patients enrolled, or the number of events (Floriani et al., 2008\nocite{Floriani}). For event-based interim analyses, group-sequential methodology is typically used (Jennison and Turnbull, 2000\nocite{Jennison}) and comes down to choosing an alpha-spending function (e.g., O’Brien-Fleming type or Pocock type) and an information fraction which play a similar role as the significance level and the sample size in power calculations for designs with only one test (Lan et al., 1994\nocite{Lan}). 

\begin{figure}[h]
\vspace*{-4mm}
\begin{center}
\includegraphics[width=0.4\textwidth]{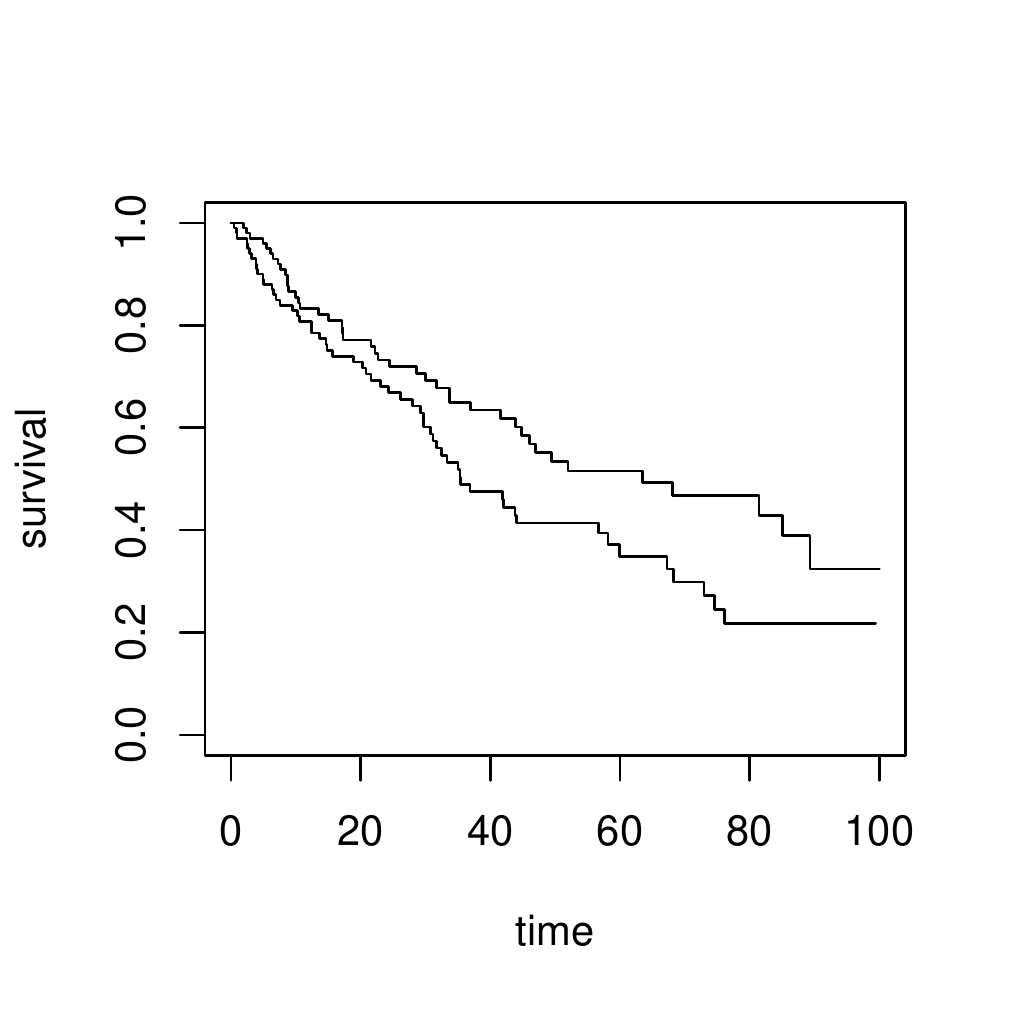} \quad
\includegraphics[width=0.4\textwidth]{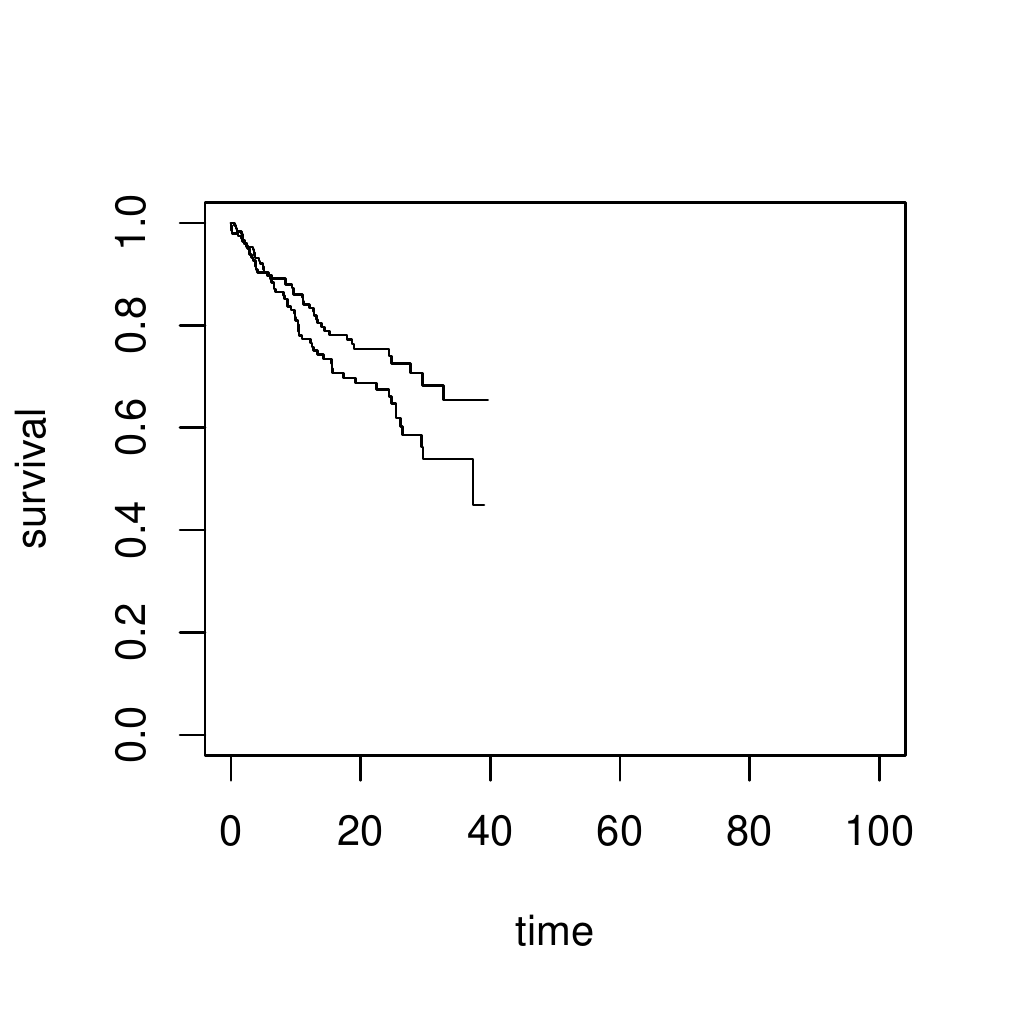}
\vspace*{-8mm}
\caption{{\footnotesize{Kaplan-Meier curves at time of the interim analysis (after 100 observed events) for a trial with 200 patients (left) and 1000 patients (right). In both trials the true survival distributions in arm $A$ and $B$ are exponential with mean 51.9 and 82.4.}}}
\label{Fig:sim}
\end{center}
\end{figure}

The amount of statistical information available at a time-point comes down to the number of observed events or the information fraction at that moment, where the latter is defined as the number of observed events divided by the total number of events planned for the final analysis. The interim analysis is performed once the information fraction equals a pre-specified value. This value can be chosen freely and should be determined in the design stage. 
However, the information fraction determines only the power of the log-rank/Cox regression test and the survival data at the interim analysis do not have to be mature in terms of Kaplan-Meier curves. Here "data maturity" is meant as in the European Medicines Agency guideline  for oncology (EMA\nocite{EMA}): "the distribution of events over time (early – late) makes it feasible to estimate the treatment effect in the full study population". As an example, to detect a hazard ratio of 0.63 in a 1:1 randomized controlled trial with 80\% power at a two-sided significance level of 0.05, 147 events are needed (Schoenfeld, 1983\nocite{Schoenfeld}). The interim analysis could be planned at an information fraction of 68\%, i.e., 100 events. 
If the trial has 200 patients much more of the Kaplan-Meier curves in each arm will be observed than when the trial has 1000 patients. This can be clearly seen in Figure \ref{Fig:sim} where the Kaplan-Meier curves have been plotted (based on simulated data) at the time of the interim analysis for the trial with 200 patients (left) and for the trial with 1000 patients (right).
More generally, if the fraction of \textsl{patients} $p$ with an event at the interim analysis is low, it might happen that the number of events at the interim analysis is sufficient for the statistical test to be statistically significant and to conclude that the new treatment is more effective than the placebo, while the survival curves are still close to $100\%$, say both above $80\%$. In that case, this conclusion of a statistical significant difference is based merely on patients who die early after the start of treatment; possibly the patients with poor prognosis. Then, it would not be sensible to claim efficacy for the whole population based on the conclusions from the interim analysis. So, when planning the moment the interim analysis will take place, it is important to have a rough idea about the survival at that moment, so that the interim analyses will not be too early ``in the survival curves''. This motivated the main question in our paper: 
``Can we derive a formula for the prediction interval for the survival rate at the time of interim analysis that 
can be used before the trial starts?''. This is not a trivial question as the time of the interim analysis is a random moment in calendar time and possibly not all patients have entered the study before the time point of the interim analysis. 

To investigate what is known regarding this question, we performed a title-abstract search in Pubmed. As there are no words that enable a direct search on the timing of an interim, we decided on a broad search using the words “interim analysis”, “interim analyses”, “interim look”, or “interim looks” in Biometrical Journal, Biometrics, Biometrika, Clinical Trials,  Contemporary Clinical Trials, Controlled Clinical Trials, Journal of the American Statistical Association, Lifetime Data Analysis, Pharmaceutical Statistics, Statistical Methods in Medical Research, and Statistics in Medicine (458 hits on 22 April 2022). Most of the literature found dealt with critical values for testing at interim analyses in specific situations (e.g.\ if the hypotheses relate to a survival probability at a fixed follow-up time (Lin et al., 1996\nocite{Lin}), Bayesian adaptive designs, estimation and bias correction, treatment arm selection, optimal designs, and sample size adaption). Papers treating questions related to the timing of an interim were found. For example: when the interim analysis based on number of events occurs in calendar time (Bagiella and Heitjan, 2001\nocite{Bagiella}) or conversely, how to update predictions regarding the number of events and thus power at fixed (non-stochastic) calendar times (Royston and Barthel, 2010\nocite{Royston2010});  the problem how to translate the calendar-time scale into information-time  scale, i.e., the fraction of events (Lan and Lachin, 1990\nocite{Lan1990}); or estimation of percentiles at a given calendar time (safety for medical devices (Murray et al., 2013\nocite{Murray})). We found one concrete advice in Van Houwelingen \textsl{et al.} (2005)\nocite{VanHouwelingen}: they advise in case of non-proportional hazards to only plan an interim analysis if a final time horizon for the final analysis is specified and at the time of the interim analysis sufficient information is present over the whole time interval up to that horizon. However, their arguments and results do not involve the Kaplan-Meier estimates at the interim analysis. In conclusion, it seems that our question has not been addressed in the literature. 

Therefore, in this paper we will consider the following situation. Suppose the interim analysis is conducted once $p 100\%$ of the patients has had an event. We will derive 
intervals that will contain the Kaplan-Meier curves at the moment of the interim analysis with high probability. The boundaries of these intervals are dependent of the expected shape of the survival curves in both arms, the accrual rate, and the fraction of patients with an event ($p$). Although the accrual rate can be influenced by opening more study sites, it is generally a given logistical restraint. Therefore, only the fraction $p$ can be freely chosen by the researcher in the design stage. By knowing the relationship between $p$ and the boundaries of the prediction interval during the design phase of the trial, choices for the fraction $p$ can be made to be sure that the  Kaplan-Meier curves at the time of the interim analysis are far enough below 100\%. The boundaries of the prediction intervals are derived from the asymptotic distribution of the Kaplan-Meier curves at the random time point the interim analysis takes place. 


The paper is outlined as follows. In Section 2 the research aims are made specific by describing them in mathematical terms. In order to do this, notation will be introduced. Next, in Section 3 the asymptotic distributions of the Kaplan-Meier curves at the time of the interim analysis are given in a two-arm and a single arm study. The proofs of these theorems are given in the Appendix. Then, in Section 4, the results of simulation studies for a range of settings are described to confirm that the asymptotic theory can be used for finite samples. Results are tabulated, to make the theory directly available for planning and R-code is provided in the appendix. Further, in Section 5, it is explained how the asymptotic theory can be applied in practice when planning a trial. The paper ends with a discussion in Section 6.

\section{Notation and specific aim}
\subsubsection*{Notation}
We consider a  clinical trial with two arms ($A$ and $B$), with a survival outcome and an interim analysis. Two time lines are important: the follow-up and the calendar time. In survival analysis, the follow-up time is usually the time line researchers are interested in as it describes the time to an event of interest from a pre-specified starting point, for instance start of treatment. The calendar time line starts at the moment the trial is started (time zero) until it is stopped. This time line is important since the moment the interim analysis is performed is defined in calendar time. For  estimating the survival curve at the interim analysis only information that is available at that moment can be used.

In a clinical trial, the effect of a new treatment is studied by comparing a survival outcome (e.g., overall survival) between the arms $A$ and $B$. To distinguish between the two arms in the notation of the observations and their distribution functions a subscript $A$ or $B$ is used. 
For a patient in arm $A$, define $T_A$ and $C_A$ as the time from entering the study to the event of interest (the survival time) and censoring, both in follow-up time. It is assumed that $T_A$ and $C_A$ are stochastically independent. Denote the distribution of the survival time $T_A$ by $F_A$, with continuous density $f_A$, and survival function $S_A=1-F_A$. The hazard function for $T_A$ is defined as $\lambda_A(t)=f_A(t)/(1-F_A(t))=f_A(t)/S_A(t)$, and its corresponding cumulative hazard function as $\Lambda_A(t) = \int_0^t\lambda_A(s)\rd s$.
Similar notation is used for arm $B$.
In calendar time the start of the study is time zero. Let $E_A$ be the time (since the start of the study) a patient from arm $A$ enters the study with distribution function $G_{Acc}$.  
In calendar time, the event-of-interest takes place at time $E_A+T_A$ (time since the start of the study) or the  patient is censored at time $E_A+C_A$, whichever comes first. For $L$ the moment (in calendar time) the study is temporary stopped (in case of an interim analysis) or definitely ended, $E_A+C_A=L$ by definition ($L$ will be specified later on). Here  and later on, we will use phrases like “temporarily stopped” and “the time of the interim analysis” interchangeably, because from an analysis point of view, patients recruited after or events occurring after the interim analysis do not play a role in the analysis. From a trial logistics point of view, recruitment and follow-up typically continue (unless there is a safety concern that has to be sorted out). 

The number of patients in the arms $A$ and $B$ are $n$ and $m$, respectively. The observations for patient $i$ in arm $A$ are given by $(E_{A,i}, T_{A,i}\wedge C_{A,i},\Delta_{A,i})$ if $E_{A,i}<L$, where  $T_{A,i}\wedge C_{A,i} = \min\{T_{A,i},C_{A,i}\}$ and $\Delta_{A,i}=1\{T_{A_i}\leq C_{A,i}\}$ equals 1 if $T_{A,i}\leq C_{A,i}$ and 0 otherwise. If $E_{A,i}>L$, patient $i$ did not enter the study before it was stopped (temporary) at time $L$ and there are no observations available. The observations of different patients are assumed to be independent. A similar notation is used in arm $B$.

\subsubsection*{Specific aim}
\begin{figure}[h]
\begin{center}
\includegraphics[width=0.45\textwidth]{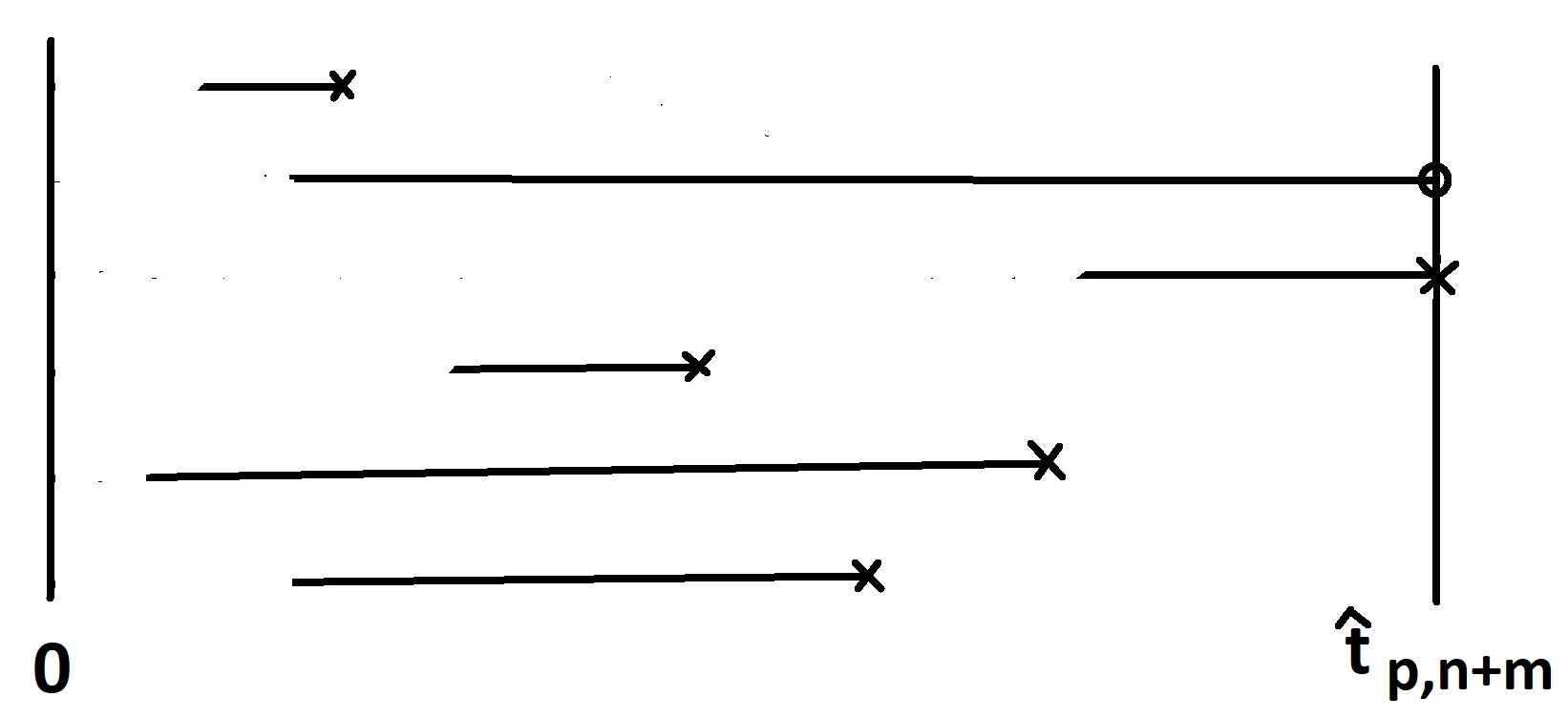} \quad
\includegraphics[width=0.44\textwidth]{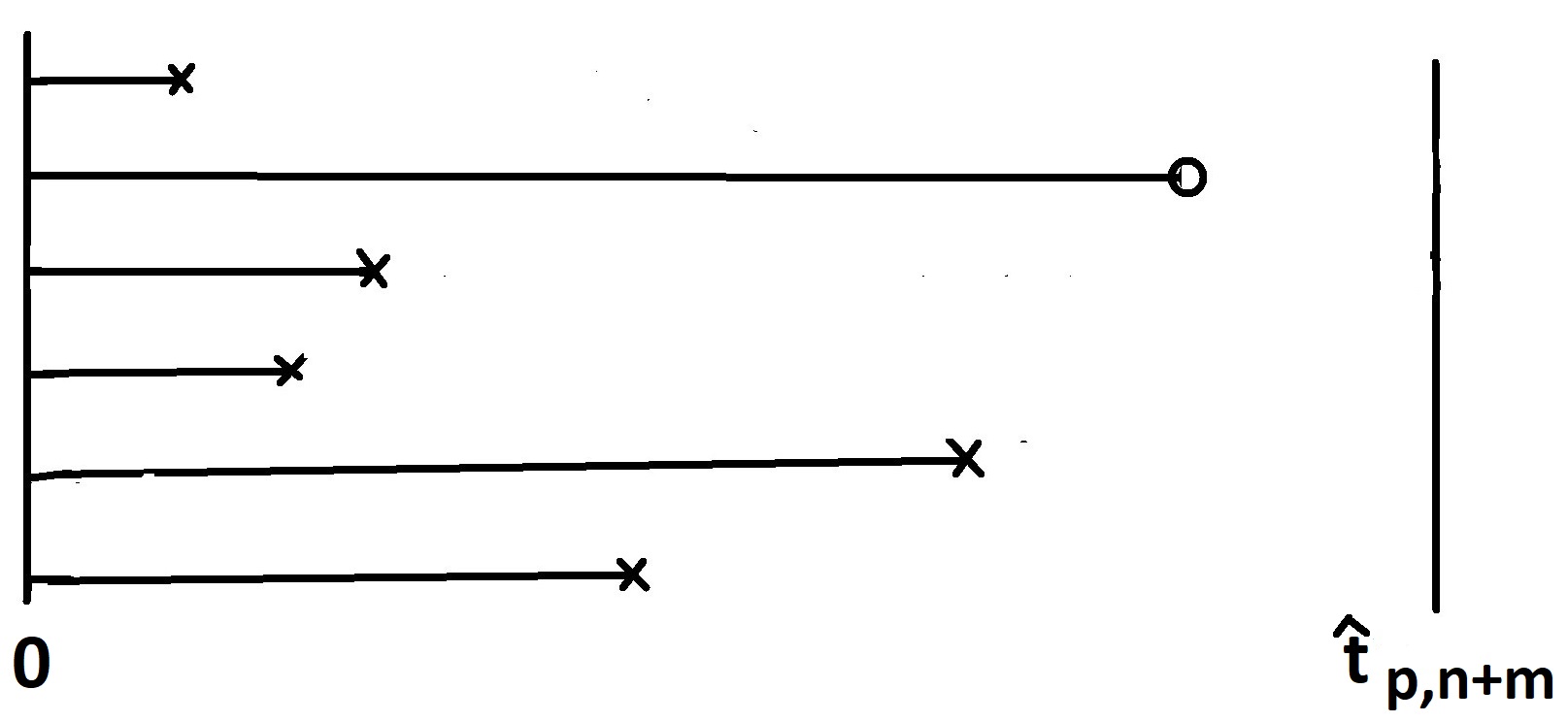}
\caption{{\footnotesize{Six patients enter the study at a random moment before the interim analysis. Left plot: six patients in calendar time. Right plot: the same six patients in follow-up time. An event of interest is represented as a cross, censoring as a circle. Although, some of the patients are at risk during the interim analysis (so in calendar time), they do not have a follow-up time of  $\hat t_{p,n+m}$.}}}
\label{fig:EnterDuringStudy}
\end{center}
\end{figure}

As was illustrated in the introduction, the Kaplan-Meier curve at the time of the interim analysis will not depend on the (statistical) information fraction, but rather on the fraction of patients  with an event. Therefore, suppose the interim analysis is performed once $100p\%$ of the patients has had an event, no matter whether these patients are from arm $A$ or arm $B$. The fraction $p$ is a direct consequence of the number of events (typically chosen based on power considerations) and the total number of patients (typically chosen on logistic feasibility). Note that the number of events and number of patients are chosen during the design stage of the trial, whence one can steer these in the design stage towards a value of $p$ which gives meaningful data maturity at the time of the interim analysis. The stochastic moment of the interim analysis is denoted as $\hat t_{p,n+m}$, and is a moment in calendar time. At the interim analysis, only observations up to that moment can be used for estimation. Specifically, only if a patient enters the study before the moment the interim analysis takes place the patient will be included in the analysis. In Figure \ref{fig:EnterDuringStudy} the follow-up time for six patients who entered the study before the interim analysis is shown. In the left plot, the events  are given in calendar time. In the right plot, the follow-up times of these patients are shown. The time-point $\hat t_{p,n+m}$ is given as well. From the figure it is immediately clear that none of the patients can have a follow-up time of at least $\hat t_{p,n+m}$ (almost surely). That means that estimates of the survival curves $S_A$ and $S_B$ at the (stochastic) point $\hat t_{p,n+m}$ (in follow-up time) are unreliable even if the sample size is high. Therefore, the aim is to consider the Kaplan-Meier curves in both arms at the time point $\hat t_{p,n+m}-\delta$ for $\delta>0$ a pre-specified value. When designing the study different values of $\delta$ can be considered. If the sample sizes $n$ and $m$ increase to infinity, $\hat t_{p,n+m}$ converges in probability to a value $t_p$ defined as the $p$th quantile of a mixture of  distributions (that will be defined later) and for large, but finite $n$ and $m$ a positive fraction of the patients will have a follow-up time that is larger than $\hat t_{p,n+m}-\delta$ under the assumption that this mixture distribution is strictly increasing, at least in a neighborhood of $t_p$.

In Section 3 (and the appendix) it will be proven that $\sqrt{n}(\tilde S_{A,n}(\hat t_{p,n+m}-\delta)-S_A(t_p-\delta))$ is asymptotically normal with mean zero and a variance $\sigma_{A,\delta}^2$, with $\tilde S_{A,n}$ the Kaplan-Meier estimator in arm $A$ based on observations up to the interim analysis at time $\hat t_{p,n+m}$. The asymptotic variance $\sigma_{A,\delta}^2$ is a function of the survival functions $S_A, S_B$, of $G_{Acc}$ (for the definition of $G_{Acc}$, see the beginning of Section 2)  and of the fraction $p$ (see Section 3 for the expression of $\sigma_{A,\delta}$). From this asymptotic behavior it follows that  
\begin{align*}
\rP\Big(S_A(t_{p}-\delta) - \xi_{\alpha/2}\frac{\sigma_{A,\delta}}{\sqrt{n}} \; \leq \; \tilde S_{A,n}(\hat t_{p,n+m}-\delta) \; \leq \; S_A(t_{p}-\delta) + \xi_{\alpha/2}\frac{\sigma_{A,\delta}}{\sqrt{n}}\Big) \approx 1- \alpha,
\end{align*} 
for $\xi_{\alpha/2}$ the upper $\alpha/2$-quantile of the standard normal distribution. Therefore, the probability $\tilde S_{A,n}(\hat t_{p,n+m}-\delta)$ will be in the interval 
\begin{align}
\Big[S_A(t_{p}-\delta) - \xi_{\alpha/2}\frac{\sigma_{A,\delta}}{\sqrt{n}}\; ; \; S_A(t_{p}-\delta) + \xi_{\alpha/2}\frac{\sigma_{A,\delta}}{\sqrt{n}}\Big]
\label{interval}
\end{align}
is approximately equal to $1-\alpha$. The boundaries of the interval depend on the true survival functions $S_A, S_B$, of $G_{Acc}$ and of the fraction $p$ (via $\sigma_{A,\delta}$). The same boundaries hold for the Breslow estimator $\hat S_{A,n}$. For the survival curves in arm $B$ a similar interval can be constructed.
 
When planning a trial, the chosen distribution functions for $S_A, S_B$ and $G_{Acc}$ play an important role for determining the sample size. The value $t_p$ is a function of, among others, these curves as well and $S_A(t_p-\delta)$ and $S_B(t_p-\delta)$ can be computed when designing the study. That means that in the designing stage of the trial $p$ (and $t_p$) can be chosen so that $S_A(t_p-\delta)$ and $S_B(t_p-\delta)$ and the corresponding boundaries are sufficiently below 1 so that the  survival curves are sufficiently mature for the aim at hand. In Section \ref{sub:Example} it is explained how to use this interval in a practical setting.\\

\section{Asymptotic results}
\label{AsymptoticResults}
In this section the asymptotic distribution of the Kaplan-Meier and the Breslow estimators evaluated at the stochastic time $\hat t_{p,n+m}-\delta$ (in follow-up time) and based on all observations that occurred before the moment of the interim analysis at $\hat t_{p,n+m}$ (in calendar time) in a two arm clinical trial (Theorem 1) and a single arm trial (Theorem 2) is given. The proofs of the theorems are given in the Appendix. Notation that is used in the theorem, especially in the definition of the asymptotic variance $\sigma^2_{A,\delta}$, is given below the theorem. \\ 

\bigskip

\noindent
{\bf{Theorem 1: Clinical trial with two arms}}\\
Suppose the interim analysis is performed once $100p\%$ of the patients has had an event, irrespective of the arm in which the events took place, at time-point $\hat t_{p,n+m}$. Let $\hat S_{A,n}$ and $\tilde S_{A,n}$ be the Breslow and the Kaplan-Meier estimators for $S_A$. Let $\delta>0$ be pre-specified, $\hat t_{p,n+m}^\delta=\hat t_{p,n+m}-\delta$ and its limit $t_p^\delta=t_p-\delta$. The asymptotic distribution of the Breslow estimator is given by
\begin{align*}
\sqrt{n}(\hat S_{A,n}(\hat t_{p,n+m}^\delta)-S_A(t_p^\delta)) \leadsto {\cal{N}}(0,\sigma_{A,\delta}^2),
\end{align*}
and of the Kaplan-Meier estimator the asymptotic distribution is the same:
\begin{align*}
\sqrt{n}(\tilde S_{A,n}(\hat t_{p,n+m}^\delta)-S_A(t_p^\delta)) \leadsto {\cal{N}}(0,\sigma_{A,\delta}^2),
\end{align*}
(where $\leadsto$ is the notation for convergence in distribution) as $n,m\rightarrow \infty$ and
\begin{align*}
\sigma_{A,\delta}^2 &= S_A(t_p^\delta)^2 \int_0^{t_p^\delta} \frac{\rd\Lambda_A(s)}{1-H_{A,t_p}(s)} \; + \; \frac{q_A f_A(t_p^\delta)^2}{(h_{mix,t_p}^{uc,\star}(t_p))^2}\; p(1-p) \\[4pt]
&\qquad -2S_A(t_p^\delta)\frac{q_A\sqrt{q_A}f_A(t_p^\delta)}{h^{uc,\star}_{mix,t_p}(t_p)}\bigg((1 - H_{A,t_p}^{uc,\star}(t_p))\Lambda_A(t_p^\delta)\;+\; \int_0^{t_p^\delta} \frac{(H_{A,t_p}^{uc}(s)-H_{A,t_p}^{uc,\star}(t_p)H_{A,t_p}(s)) \rd\Lambda_A(s)}{1-H_{A,t_p}(s)}\bigg).
\end{align*}
\noindent

\bigskip

The proof of this theorem is given in Appendix B. For the Breslow estimators $\hat S_{B,m}$ and the Kaplan-Meier estimator $\tilde S_{B,m}$ in arm $B$, a similar result holds. The variance $\sigma^2_{A,\delta}$ depends on multiple distribution functions and parameters. The notation will be explained below. As in calendar time only observation up to the interim analysis are used, asymptotically $L=t_p$, and  the observations are censored in calendar time at time $t_p$: so $E_A+C_A=t_p$.
The (sub)distribution functions $H_{A,t_p}$ and $H^{uc}_{A,t_p}$ are defined in follow-up time as   $H_{A,t_p}(t)=\rP(T_A\wedge C_A \leq t)$ and $H^{uc}_{A,t_p}(t) = \rP(T_A\wedge C_A \leq t,\Delta_A=1) = \rP(T_A \leq t,\Delta_A=1)$, where the latter one is for uncensored observations (the superscript ``uc'' stands for ``uncensored''). With similar notation for arm $B$, 
the mixture of the distributions in the two arms is defined as
\begin{align*}
H_{mix,t_p}(t) = q_AH_{A,t_p}(t) + q_BH_{B,t_p}(t),\qquad 
H^{uc}_{mix,t_p}(t) = q_AH^{uc}_{A,t_p}(t) + q_BH^{uc}_{B,t_p}(t), 
\end{align*}
with $q_A=\lim_{n,m\to \infty} \; n/(n+m)$ and $q_B= \lim_{n,m\to \infty}\; m/(n+m)$. In calendar time the definitions are very similar, define the (sub-) distribution functions  $H^{\star}_{A,t_p}(t) = \rP(E_A+(T_A\wedge C_A) \leq t)$ and $H^{uc,\star}_{A,t_p}(t) = \rP(E_A+T_A\leq t,\Delta_A=1)$.
Their densities are denoted as $h^{\star}_{A,t_p}$ and $h^{uc,\star}_{A,t_p}$, respectively. The mixtures of the distributions in the two arms are defined as $H^{\star}_{mix,t_p}(t) = \; q_A H^{\star}_{A,t_p}(t) \;+\; q_B H^{\star}_{B,t_p}(t)$ and $H^{uc,\star}_{mix,t_p}(t) = \; q_A H^{uc,\star}_{A,t_p}(t) \;+\; q_B H^{uc, \star}_{B,t_p}(t)$.  
Remember that the interim analysis is performed once $100 p\%$ of all patients has had an event; at time point $\hat t_{p,n+m}$. This stochastic time point $\hat t_{p,n+m}$ converges in probability to $t_p$ defined as the $p$th quantile of the mixture $H_{mix,t_p}^{uc,\star}$.


\bigskip

In the next theorem the single arm setting is considered. Since we do not have to distinguish between arms, the notation $A$ and $B$ is left-out from the notation. 

\bigskip

\noindent
{\bf{Theorem 2: Clinical trial with single arm}}\\
Define $\hat t_{p,n}$ as the moment $100p\%$ of the patients had an event. Let $\hat S_n$ and $\tilde S_n$ be the Breslow and the Kaplan-Meier estimators based on the observations up to time $\hat t_{p,n}$. Let $\delta>0$ be fixed, $\hat t_{p,n}^\delta=\hat t_{p,n}-\delta$ and its limit $t_p^\delta=t_p-\delta$. For the Breslow-estimator it holds that 
\begin{align*}
\sqrt{n}(\hat S_n(\hat t_{p,n}^{\delta})-S(t_p^{\delta})) \; \leadsto\; {\cal{N}}(0,\sigma^2_\delta),
\end{align*}
and for the Kaplan-Meier estimator that
\begin{align*}
\sqrt{n}(\tilde S_n(\hat t_{p,n}^{\delta})-S(t_p^{\delta})) \; \leadsto\; {\cal{N}}(0,\sigma^2_\delta),
\end{align*}
as $n\rightarrow \infty$, with  
\setlength{\jot}{10pt}
\begin{align}
\sigma^2_\delta &= S(t_p^{\delta})^2 \int_0^{t_p^{\delta}} \frac{\rd\Lambda(s)}{1-H_{t_p}(s)} \; + \; \frac{f(t_p^{\delta})^2}{(h^{uc,\star}_{t_p}(t_p))^2} \; p(1-p) \nonumber \\[4pt]
& \qquad -2S(t_p^{\delta})\frac{f(t_p^{\delta})}{h^{uc,\star}_{t_p}(t_p)}\bigg((1 - p)\Lambda(t_p^{\delta}) + \int_0^{t_p^{\delta}} \frac{(H^{uc}_{t_p}(s)-H^{uc,\star}_{t_p}(t_p)H_{t_p}(s)) \rd\Lambda(s)}{1-H_{t_p}(s)}\bigg).
\label{VarOneArm}
\end{align}

\bigskip

\noindent
Theorem 2 follows from Theorem 1 by taking $m=0$ and, thus, $q_A=1$ and $q_B=0$.

\medskip

The asymptotic variance in (\ref{VarOneArm}) is a sum of three terms. The first term can be seen as the variance due to the estimation of the survival function $S$, the second term due to the estimation of the time point $t_p$ and the last term comes from the covariance between the two terms. This covariance is negative as $\hat S_n$ and $\hat t_{p,n}$ are negatively correlated. If the sum of the second and third term in the display is negative, the estimator $\hat S_{n}(\hat t_{p,n})$ for estimating $S(t_p)$ has an asymptotically smaller variance than the estimator $\hat S_{n}(t_p)$, even though the second estimator is determined at a fixed time point $t_p$. This is not surprising, because of the following example. Consider the situation in which there is no censoring. In that case the Kaplan-Meier curve equals the empirical survival curve and $\tilde S_n(\hat t_{p,n})$ equals $1-p$ by definition and the asymptotic variance will be equal to zero. The asymptotic variance of $\tilde S_{n}(t_p)$ equals $p(1-p)$ which is larger than zero.

\section{Simulation studies}
\subsection*{Comparison of asymptotic versus simulation results}
In Theorem 1 the asymptotic distributions of the Kaplan-Meier and the Breslow estimators are given in a two arm trial. Based on this asymptotic distribution it follows that the probability $\tilde S_{A,n}(\hat t_{p,n+m})$ will be in the interval  
\begin{align*}
\Big[S_A(t_{p}^\delta) - \xi_{\alpha/2}\frac{\sigma_{A,\delta}}{\sqrt{n}}\; ; \; S_A(t_{p}^\delta) + \xi_{\alpha/2}\frac{\sigma_{A,\delta}}{\sqrt{n}}\Big]
\end{align*}
is approximately equal to $1-\alpha$. In this subsection the aim is to study the accuracy of the asymptotic interval by comparing it to the interval obtained by Monte Carlo simulations. 

In all scenarios it is assumed that in both arms the time to the event of interest follows an exponential distribution.
In total eight different settings are considered, obtained by varying the hazard ratio for the two arms, the severity of the disease (in terms of the median survival time) and the rarity of the disease (in terms of the accrual period that is necessary to include the patients). 
More specifically, we consider
\begin{itemize}
\item Effect of the treatment in terms of the hazard ratio: we consider two situations:
\vspace{-0.4em}
\begin{itemize}
\item Strong effect: hazard ratio equals 0.65. This implies that $\theta_B / \theta_A = 0.65$.
\item Median effect: hazard ratio equals 0.75. This implies that $\theta_B / \theta_A = 0.75$.
\end{itemize}
\item Severity of the disease:
\vspace{-0.4em}
\begin{itemize}
\item Aggressive: the median survival time is 6 months; $\theta_A= -\log(0.5)/6=0.12$.
\item Indolent: the median survival time is 36 months; $\theta_A= -\log(0.5)/36=0.019$.
\end{itemize}
\item Rarity of the disease:
\vspace{-0.4em}
\begin{itemize}
\item Rare: accrual is 4 patients per month; the accrual time equals $(n+m)/4$ months.
\item Frequent: accrual is 20 patients per month; the accrual time equals $(n+m)/20$ months.
\end{itemize}
\end{itemize}

In a clinical trial the sample size is usually determined to have sufficient power for the log rank test at the end of the study. In fact, what counts is that the number of patients together with the follow-up ensures a sufficient (expected) number of events. The required expected number of events for a log-rank test (Schoenfeld, 1991\nocite{Schoenfeld1981}) or Cox regression (Schoenfeld, 1983\nocite{Schoenfeld}) is calculated via the Schoenfeld’s formula: $d=\#\mbox{events} = (\xi_{\alpha/2}+\xi_{\beta})^2/((\log \mbox{HR})^2 q_A q_B)$, for HR the hazard ratio. In case of a 1-1 randomisation, $\alpha=0.05$ (two-sided) and 80\% power, $\# \mbox{events} = 31.4/(\log \mbox{HR})^2$. A common approach is to choose a combination of number of patients $n+m$, accrual distribution function $G_{Acc}$, and follow-up duration after the recruitment of the last patient, $FU$, such that the expected number of events (after the last recruited  patient has $FU$ follow-up) equals the required expected number of events $d$. If the recruitment is uniform over a recruitment period from $0$ to $R$ in calender time (i.e., $G_{Acc}$ has density $1/R$ on the interval from $0$ to $R$ and equals $0$ elsewhere), then the duration of the trial is $L=R+FU$ and the expected number of events in the period from $0$ to $L$ is in arm $A$:
\begin{align}
n\; H^{uc,\star}_{A,L}(L) & 
=n\; \rP(T_A \leq L -E_A \, ,\, \Delta_A=1)
=n/R \; \int_0^R \rP( T_A \leq L-s) \,\rd s  \nonumber
\\
&=n/R \int_0^R \Big( 1- \exp(-\lambda_A(L-s)) \Big) \, \rd s =
n\; \Big[ 1- \frac{ \exp(-\lambda_A L) }{\lambda_A R} \; \big(\exp(\lambda_A R)-1\big)   \Big]  \,, 	
\end{align}
and a similar expression holds for arm $B$. The combination of sample size, recruitment time $R$, and follow-up $FU$ is chosen such that $n\; H^{uc,\star}_{A,L}(L) + m \; H^{uc,\star}_{B,L}(L)=d$. 

In order to reduce the number of different settings, the following situation is considered. First, randomization is 1:1: $n=m$ in the formula. Also, the follow-up time after the last patient is recruited $FU$ is fixed at $6$ months. Then still there are several combinations of recruitment period $R$ and total sample size $2n$ that can provide the required number of events at the end of follow-up. Depending on the occurrence of the disease, the recruitment rates 4 or 20 patients per month are considered. The rates fix the ratio of sample size $2n$ and recruitment period $R$ and result in one combination of $2n$ and $R$. For the severity of the disease (i.e., the survival curves), aggressive and indolent diseases are considered. 
For the two scenario's for the treatment effect ($hr = 0.65$ or $hr= 0.75$), the number of events needed are equal to 170 and 380, respectively. 

The interim analysis is often performed once a fraction of the required number of events for the final analysis is observed. This fraction is called the information fraction (IF). If (and only if) the total number of patients is decided on, the IF is one-to-one related to the fraction $p$, the fraction of patients with an event at the interim analysis (the fraction $p$ could be called the patient fraction to contrast it with the information fraction).

In the simulation study we consider two moments of interim analysis:
\begin{itemize}
\item Early interim analysis: after 40\% of the events: IF$=0.40$.
\item Late interim analysis: after 60\% of the events: IF$=0.60$.
\end{itemize}
The different settings that are considered are given in Table 1. 

The asymptotic interval is found by computing its bounds based on the chosen setting. The finite sample intervals are found as follows. Data are sampled for the $n+m$ patients in the study. 
Based on the sampled data the estimate $\hat S_{A,n}(\hat t_{p,n+m}^\delta)$ is computed. This is repeated 1000 times. The sample mean and the 2.5\% and 97.5\% quantiles of the estimates  are used to construct a 95\% prediction interval.

\begin{table}[h]
\footnotesize
\begin{center}
\begin{tabular}{|c|c|c|c||c|c|c|c||c|c|c|c|}
\hline
\multicolumn{1}{|c|}{}&
\multicolumn{3}{c||}{disease/treatment}&
\multicolumn{4}{c||}{trial}&
\multicolumn{4}{c|}{interim}\\
nr. &  effect & severity  & rarity  & $n+m$ & total & accrual & FU & IF & \# events  &  $p$ & $t_p$ \\
& (HR) & (median) & (pat/mon.) & patients & \# events & (mon.) & (mon.) & & & &\\
\hline
1. &  0.65 & 6  & 4  & 196 & 170 & 49  & 6 & 0.40  & 68 &      0.35 & 27  \\ 
2. &  0.65 & 6  & 20  & 260 & 170 & 13 & 6  & 0.40  & 68 & 0.26  & 10  \\ 
3. & 0.65 & 36  & 4  & 344 & 170 & 86 & 6 & 0.40 &  68 &  0.20  & 53 \\
4. & 0.65 & 36  & 20  & 620 & 170 & 31 & 6 & 0.60 &  102   & 0.16 & 27 \\
\hline
5. & 0.75 & 6 & 20  & 480 & 380 & 24 & 6 & 0.40 &  152 & 0.32 & 16 \\
6. & 0.75 & 36 & 4  & 580 & 380 & 145 & 6 & 0.40 &  152 &  0.26 & 82  \\
7. & 0.75 & 36  & 20  & 1000 & 380 & 50 & 6 & 0.40 &  152 & 0.15 & 33 \\
8. & 0.75 & 36  & 20  & 1000 & 380 & 50 & 6 & 0.60 &  228 & 0.23 & 41 \\
\hline
\end{tabular}
\caption{{\footnotesize{Different scenario's for simulation study. In all cases $n=m$; the number of patients in the two arms  equal. The variable ``total \# events'' indicates the total number of events needed based on the Schoenfeld's formula for $\alpha=0.05$ and a power of 0.80. The variable ``\# events'' (for the interim analysis) indicates the number of observed events at time of the interim. ``FU'' is the follow-up time after the last patient was accrued. }}}
\end{center}
\label{Table:Scenario}
\end{table}

\begin{table}[h]
\footnotesize
\begin{center}
\begin{tabular}{|c||cc|cc||cc|cc|}
\hline
\multicolumn{1}{|c||}{}&
\multicolumn{4}{c||}{Arm $A$}&
\multicolumn{4}{c|}{Arm $B$}\\
nr. & $\hat S_{A,n}(\hat t_{p,n+m}^\delta)$  & 95\% int & $S_A(t_p^\delta)$ & asymp int & $\hat S_{B,m}(\hat t_{p,n+m}^\delta)$  & 95\% int & $S_B(t_p^\delta)$ & asymp int \\
\hline
1. & 0.07 & [0.00 ; 0.20] & 0.06 & [-0.06 ; 0.18] & 0.17 & [0.00 ; 0.36] & 0.16 & [-0.02 ; 0.34] \\
2. & 0.36 & [0.20 ; 0.50] & 0.36 & [0.20 ; 0.52] & 0.52 & [0.35 ; 0.66] & 0.52 & [0.36 ; 0.68] \\
3. & 0.39 & [0.23 ; 0.53] & 0.40 & [0.25 ; 0.55] & 0.55 & [0.38 ; 0.69] & 0.55 & [0.40 ; 0.70] \\
4. & 0.63 & [0.53 ; 0.70] & 0.63 & [0.54 ; 0.72] & 0.74 & [0.66 ; 0.81] & 0.74 & [0.66 ; 0.82] \\
\hline
5. & 0.20 & [0.07 ; 0.30] & 0.20 & [0.09 ; 0.31] & 0.30 & [0.16 ; 0.41] & 0.30 & [0.17 ; 0.42] \\
6. & 0.24 & [0.13 ; 0.36] & 0.24 & [0.13 ; 0.35] & 0.34 & [0.21 ; 0.46] & 0.34 & [0.22 ; 0.46] \\
7. & 0.57 & [0.48 ; 0.65] & 0.57 & [0.48 ; 0.65] & 0.65 & [0.56 ; 0.73] & 0.65 & [0.57 ; 0.73] \\
8. & 0.49 & [0.41 ; 0.56] & 0.49 & [0.41 ; 0.57] & 0.58 & [0.50 ; 0.66] & 0.58 & [0.51 ; 0.66] \\
\hline
\end{tabular}
\caption{{\footnotesize{Results of the simulation results. The number in the first column refers to the simulation scenario in Table 1. 
In all cases $\delta$ was taken equal to $\delta=0.1\; t_p$.}}}
\end{center}
\label{Table:Results}
\end{table}

The results of the simulation study (Table 2) show that the asymptotic intervals are accurate as they show a very strong resemblance to the intervals constructed based on Monte Carlo simulations. Further, in all settings the intervals are sufficiently narrow to be useful for planning an interim analysis.

\subsection*{The value $\delta$ and the width of the prediction interval}
Remember that the interim analysis is performed at time $\hat t_{p,n+m}$ in calendar time. Since no patients can have a follow-up time of at least $\hat t_{p,n+m}$ (see Figure \ref{fig:EnterDuringStudy}), the survival curves in the two arms are estimated at time $\hat t_{p,n+m}-\delta$ in follow-up time, with $\delta>0$. If $\delta$ is small, only a few patients may have a follow-up time of at least $\hat t_{p,n+m}-\delta$ and the Kaplan-Meier estimator in $\tilde t_{p,n+m}-\delta$ may be inaccurate, what will lead to a wide(r) interval for $S_A(t_p-\delta)$ and $S_B(t_p-\delta)$. At the other hand, if $\delta$ is large, the Kaplan-Meier (or Breslow) estimator is evaluated earlier in follow-up time and may not represent all the information that is available at the interim analysis. These opposing considerations are similar to a bias-variance trade-off. The value $\delta$ can be chosen by the researcher and, therefore, it is  interesting to study the effect of $\delta$ on the width of the prediction interval. The settings 3 and 7 (see Table 1) are considered for different values of $\delta$. From Table 3 it can be seen that for an increasing value of $\delta$ the value of $t_p-\delta$ becomes smaller and the survival curve in $t_p^\delta=t_p-\delta$ increases (by definition). Moreover, it can be seen that the widths of the prediction intervals decrease with increasing $\delta$. The latter is a direct consequence of the fact that the survival curves can be estimated more accurately if more patients are still at risk. More settings have been considered. The conclusions are the same (results not shown).

\begin{table}
\footnotesize
\begin{center}
\begin{tabular}{|c||cc|cc||cc|cc|}
\hline
\multicolumn{1}{|c||}{}&
\multicolumn{4}{c||}{Arm $A$}&
\multicolumn{4}{c|}{Arm $B$}\\
$\delta$ & $\hat S_{A,n}(\hat t_{p,n+m}^\delta)$  & 95\% int & $S_A(t_p^\delta)$ & asymp int & $\hat S_{B,m}(\hat t_{p,n+m}^\delta)$  & 95\% int & $S_B(t_p^\delta)$ & asymp int \\
\hline
0.01 & 0.37 & [0.00 ; 0.53] & 0.36 & [0.14 ; 0.58] & 0.51 & [0.26 ; 0.67] & 0.52 & [0.30 ; 0.73]\\
0.10 & 0.40 & [0.22 ; 0.53] & 0.40 & [0.25 ; 0.55] & 0.55 & [0.37 ; 0.69] & 0.55 & [0.40 ; 0.70]\\
0.25 & 0.47 & [0.34 ; 0.58] & 0.46 & [0.34 ; 0.59] & 0.61 & [0.50 ; 0.72] & 0.61 & [0.49 ; 0.73]\\
\hline
0.01 & 0.54 & [0.39 ; 0.64] & 0.54 & [0.41 ; 0.66] & 0.62 & [0.49 ; 0.72] & 0.63 & [0.51 ; 0.74]\\
0.10 & 0.57 & [0.48 ; 0.65] & 0.57 & [0.48 ; 0.65] & 0.65 & [0.57 ; 0.73] & 0.65 & [0.57 ; 0.73]\\
0.25 & 0.62 & [0.56 ; 0.69] & 0.62 & [0.56 ; 0.69] & 0.70 & [0.65 ; 0.76] & 0.70 & [0.64 ; 0.77]\\
\hline
\end{tabular}
\caption{{\footnotesize{Results of the simulation results. The value of $\delta$: $\delta= x \; t_p$ with $x$ the value in the first column. 
}}}
\end{center}
\end{table}

\section{In practice} 
\label{sub:Example}
A researcher who aims to design a confirmatory clinical trial with the log-rank test or Cox regression analysis is typically first interested in obtaining sufficient power. In group-sequential methodology, the power of such tests at the final or the interim analysis depends on the number of events and the critical values (alpha-spending function) at these analyses. In practice, the interim analysis is often conducted after a certain percentage of the required events that is necessary for sufficient power at the final analysis (this is called the information fraction). To illustrate that information fraction relates (only) to power, consider the following. When the power (for a certain effect size) at the final analysis is 80\% (with a two-sided significance level of 0.05) and one interim analysis is planned using an O'Brien-Fleming boundary, then the power  (for achieving a statistically significant test statistic already at that one interim analysis) is 6\% for 40\% information fraction (IF), 18\% for IF=50\%,  34\% for IF=60\%, 51\% for IF=70\%, 66\% for IF=80\%, and 76\% for IF=90\%. Conversely, 80\% power for the interim analysis can only be achieved if the true effect is larger than what  was supposed for the final analysis. The factor by which this should be larger (on log hazard ratio scale) is 2.136 for IF=40\%, 1.864 for IF=50\%, 1.554 for IF=60\%, 1.330 for IF=70\%, 1.166 for IF=80\% and 1.051 for IF=90\%. 
 
Besides power, also the maturity of the survival data at the time of the interim analysis plays a role. In the formula derived to estimate the prediction interval of the Kaplan-Meier curves at the interim analysis (Section \ref{AsymptoticResults}), the input parameters are: the survival distribution in each arm, the accrual distribution function $G_{Acc}$, the relative sample sizes in each arm, the fraction $p$, and the closeness parameter $\delta$. The choice of the distributions $S_A, S_B$  are ideally based on results from a similar (e.g., explorative) study which was performed earlier or otherwise on clinical reasoning. The accrual rate and accrual time (summarized in $G_{Acc}$), the number of patients $n$ and $m$ in each arm, and the minimum follow-up time are more in the researcher's control by the choice of recruitment sites and/or determined by logistical feasibility. 

The parameter $p$ does not have to be specified for the power analysis, but can be calculated once the total number of patients has been selected. Its value reflects  the moment the interim analysis will be performed in terms of follow-up time and consequently $p$ determines the expected survival curve up to that point. The higher the value of $p$, the later the interim analysis will be performed and the more information on the survival curves will be available. Often it is important  that enough information is available at the interim analysis and $p$ should be chosen accordingly. Sufficient information could for example  mean that the survival curve up to the interim analysis can be seen as a reasonable representation of the survival for the whole patient population. On the other hand,  the interim analysis should not be too close to the end of the trial for acting on an interim analysis to be meaningful. Different values of $\delta$ could and should be considered by the researcher. By varying $\delta$, one can trade off between being close to the interim analysis versus obtaining a precision that is meaningful for the aim at hand. 

Summarised,  we envision the following strategy to determine the timing of an interim analysis. First, power (i.e., number of events) is considered as usual. To assess maturity of the survival  curves, the expected survival rates close to the time of the interim analysis are estimated using for instance our methodology. Next, the timing can then be adjusted such that not only sufficient power but also sufficient maturity of data for the purpose at hand is to be expected at the interim analysis. It is advised to investigate this for a range of plausible survival curves for both arms. 

\subsubsection*{An application}
With the proposed methodology it is assessed  whether the preplanned interim analysis for progression-free survival (PFS) in the Keynote 204 study would a priori have been expected to give mature data. The Keynote 204 was a study investigating prembolizumab versus a control of brentuximab vedotin in patients with relapsed or refractory classical Hodgkin lymphoma. The relevant design parameters were: 300 patients randomized 1:1; 12 month accrual period; exponential survival assumed with a median progression-free survival of 5.6 months in the brentuximab vedotin arm; hazard ratio of 0.622; and 176 events planned for the efficacy (i.e., confirmatory) interim analysis (see protocol page 93 and 96 in the online Appendix of Kuruvilla et al. (2021)\nocite{Kuruvilla}). The protocol did not specify the shape of the accrual rate over time, so we will assume a uniform accrual of the 300 patients over 12 months. Also, a minimum follow-up was not specified. Therefore, we will assume that the interim analysis takes place after recruitment of all patients; this is the case for $t_p=14.87$ months. 

We now evaluate the Kaplan-Meier curves at different time points close to the expected time of the interim analysis. Starting with $\delta=0.1t_p$ (this comes down to 6.4 weeks before the expected time of the interim planning), the expected Kaplan-Meier estimates (95\%-prediction interval) are 19\% (9\%-29\%) in the brentuximab vedotin arm and 36\% (24\%-47\%) in the prembolizumab arm.  For $=0.05t_p$ (3.2 weeks before the interim), this is 17\% (6\%-29\%) in the brentuximab arm and 34\% (21\%-47\%) in the prembolizumab arm. When looking  at $=0.01t_p$ (so 4.5 days before the interim analysis), we expect to see  16\% (2\%-30\%) for brentuximab arm and 32\% (16\%-49\%) for prembolizumab arm. Thus, a substantial part of the survival curves is expected to be observed in both arms. In particular, both early and late events are expected to be seen and the planned interim analysis allows to assess the survival benefit (if any) for a large majority of the population.

\section{Summary and Discussion}
In  this paper we derived the asymptotic boundaries of the  prediction interval for the survival curve at the (random) time that a prespecified total number of events has been reached, also including the situation that this number has been reached before the planned total number of patients has been recruited. The input  parameters for the derived formula  are the supposed survival curves, the patient accrual, the prespecified number of total events, and the total number of patients. These are typical parameters used to plan a trial with a survival outcome anyway, so no new information is needed. Thus, the derived formula can be easily implemented in typical planning practice. Although we focused  on an interim analysis in a two-arm trial, the expected survival estimates in the final analysis can be estimated as well.


A typical application of the proposed theory is a phase 3 randomized clinical trial where an interim analysis is planned for early stopping for efficacy and is tested using a log-rank test or a test for the Cox model. Often the interim analysis is planned when a given number of events has occurred. This will only cover whether sufficient power is expected for the statistical test. As the test relates typically to equality of whole curves, statistical significance at an interim analysis could be due to (only) early differences in the survival curves. It will not be guaranteed that survival data are sufficiently mature for drawing conclusions on the aim of the study. For example, the European Medicines Agency oncology guideline (EMA\nocite{EMA}) states that interim analyses “should be undertaken only when available survival data provide the information needed for a proper evaluation of benefit/risk”. The same guideline states “If a clear majority of the total number of expected events in the long term has been observed and a difference has been documented, this is normally accepted as an indicator that the study is reasonably mature and that the study results will remain stable over prolonged follow-up.” Of interest is that the wording “the total number of expected events” relates to the long-run (large follow-up times), not the total number of events at the final analysis. Despite this, maturity of the survival curves at the time of the interim analysis is often not considered explicitly in the planning stage or only qualitatively. Our results allow to plan for mature survival data \textsl{quantitatively}. They also give insight in the relation between the patient fraction $p$ (that is chosen by the researcher) and the amount of information that is expected to be available at the moment of the interim analysis: the Kaplan-Meier survival estimates and their corresponding prediction interval. Thus the results in this paper can help designing the trial.

In the simulation studies, the survival times are assumed to come from an exponential distribution and the accrual times from a uniform distribution. This is the most common setting that is assumed when designing a trial with survival outcome. However, the asymptotic distribution was derived for an arbitrary distribution. The R-code is available from the corresponding author upon request and can be easily adapted for other distributions (e.g., Weibul or Parmar-Royston models (Royston and Parmar, 2002\nocite{Royston2002})) to make the application more general.

\bigskip

\noindent
{\bf {CONFLICT OF INTEREST}}\\
The authors declare no potential conflict of interest.

\bigskip

\noindent
{\bf {DATA AVAILABILITY STATEMENT}}\\
No real life data have been used in this publication.

\bigskip

\noindent
{\bf ORCID}\\
Marianne Jonker: https://orcid.org/ 0000-0003-0134-8482

\newpage
\section*{Appendices: derivation of asymptotic distributions}

In Appendix B the proof of Theorem 1 is given. The proof relies on the derivation of the asymptotic distribution of the Breslow and the Kaplan-Meier estimator (based on all observations) in a random time point. The derivation of this asymptotic distribution is given in Appendix A. Below, first new notation is introduced.

\subsection*{Notation}
Suppose a clinical trial with two arms, $A$ and $B$, is performed.  Let $T_A$ be the time from entering the study to the time of the event of interest for an arbitrary patient in arm $A$. Its distribution function is denoted as $F_A$, its survival function as $S_A=1-F_A$, density as $f_A$ and its hazard function as $\lambda_A(t)=f_A(t)/(1-F_A(t))=f_A(t)/S_A(t)$. The distribution for the censoring time $C_A$ (time from entering the study to the end of the study) is given by $G_L$, for both arms, depending on $L$, where $L$ is the time the trial is temporary stopped or definitely ended (in calendar time). The distribution of the accrual time $E_A$ (in calendar time) is denoted as $G_{Acc}$ and the relationship between $G_{Acc}$ and $G_L$ is given by:
\begin{align*}
G_L(c) \;=\; \rP(C_A \leq c)\;=\;1-G_{Acc}(L-c).
\end{align*}
where we use that $E_A + C_A = L$.

Furthermore, define $H_{A,L}(t)=\rP(T_A\wedge C_A \leq t)$ and $H^{uc}_{A,L}(t) = \rP(T_A\wedge C_A \leq t,\Delta_A=1) = \rP(T_A \leq t,\Delta_A=1)$. With this notation the cumulative hazard function of $T_A$ can be written as 
\begin{align}
\Lambda_A(t) &=\; \int_0^t\lambda_A(s)\rd s \;=\; \int_0^t \frac{1}{S_A(u)}\rd F_A(u) \nonumber \\
\addlinespace
&=\; \int_0^t \frac{1-G_L(u)}{(1-F_A(u))(1-G_L(u))}\rd F_A(u) \;=\; \int_0^t \frac{1}{1-H_{A,L}(u)}\rd H_{A,L}^{uc}(u),
\label{Lambda}
\end{align}
where the last equality follows from $1-H_{A,L}(t)=\rP(T_A\wedge C_A > t) = \rP(T_A> t,C_A> t)=(1-F_A(t))(1-G_L(t))$ and the fact that $H_{A,L}^{uc}(t)=\int_0^t 1-G_L(u) \rd F_A(u)$. The survival function $S_A$ is related to the cumulative hazard function $\Lambda_A$ via the relationship
\begin{align}
S_A(t) \; = \; \exp(-\Lambda_A(t)).
\label{Ftp}
\end{align}

Consider the standard right censoring scenario with $n$ iid sampled patients. The (follow-up) observations equal $(T_{A,1}\wedge C_{A,1},\Delta_{A,1}), \ldots, (T_{A,n}\wedge C_{A,n},\Delta_{A,n})$. Based on these observations the distribution $H_{A,L}$ and the subdistribution $H^{uc}_{A,L}$ are estimated by their empirical distribution functions $\mathbb{H}_{A,n}$ and $\mathbb{H}^{uc}_{A,n}$ defined as:
\begin{align}
\label{EstHuc}
\mathbb{H}_{A,n}(t) &= \frac{1}{n}\sum_{i=1}^n 1\{T_{A,i}\wedge C_{A,i} \leq t\},\\
\mathbb{H}^{uc}_{A,n}(t) &= \frac{1}{n}\sum_{i=1}^n  1\{T_{A,i}\wedge C_{A,i} \leq t,\Delta_{A,i}=1\}= \frac{1}{n}\sum_{i=1}^n \Delta_{A,i} 1\{T_{A,i}\wedge C_{A,i} \leq t\}.
\nonumber
\end{align}

By estimating the cumulative hazard function $\Lambda_A$ by the Nelson-Aalen estimator obtained by replacing $H_{A,L}$ and $H^{uc}_{A,L}$ in (\ref{Lambda}) by their empirical distribution functions $\mathbb H_{A,n}$ and $\mathbb H^{uc}_{A,n}$,
\begin{align}
\hat\Lambda_{A,n}(t) = \int_0^t \frac{1}{1-\mathbb{H}_{A,n}(s)}\rd \mathbb{H}_{A,n}^{uc}(s),
\label{NA-estimator}
\end{align}
the Breslow estimator (Breslow, 1972\nocite{Breslow}) for $S_A$ is found:
\begin{align}
\hat S_{A,n}(t) \; = \; \exp(-\hat \Lambda_{A,n}(t))
\; = \; \exp\Big(-\int_0^t\frac{1}{1-\mathbb H_{A,n}(u)} \; \rd \mathbb H^{uc}_{A,n}(u) \Big).
\label{Stilde}
\end{align}
The distribution function $F_A$ is estimated as $\hat F_{A,n} := 1-\hat S_{A,n}$.

The survival curve $S_A$ is also related to $\Lambda_A$ via the product integral
\begin{align*}
S_A(t) \; = \; \prod_{u\leq t}(1-\rd \Lambda_A(u)),
\end{align*}
where the $\prod$-sign is used as the product integral. By replacing the cumulative hazard $\Lambda_A$ by the Nelson-Aalen estimator $\hat\Lambda_{A,n}$ the Kaplan-Meier estimator for $S_A$ is obtained:
\begin{align}
\tilde S_{A,n}(t) \; = \; \prod_{u\leq t}(1-\rd \hat\Lambda_{A,n}(u)),
\label{SKM}
\end{align}
and the corresponding estimator for $F_A$ is denoted as $\tilde F_{A,n} = 1 - \tilde S_{A,n}$.
The Breslow estimator $\hat S_{A,n}$ and the Kaplan-Meier estimator $\tilde S_{A,n}$ do not coincide exactly, but are asymptotically equivalent (Therneau and Grambsch, 2000\nocite{Therneau}).

The distributions $H_{A,L}^\star(t)=\rP(E_A+(T_A\wedge C_A)\leq t), H_{A,L}^{uc,\star}(t)=\rP(E_A+(T_A\wedge C_A)\leq t,\Delta_A=1)$, as well as those for arm $B$, are estimated by their corresponding empirical distribution functions; for $H_{A,L}^\star$ and $H_{A,L}^{uc,\star}$ these are denoted as $\mathbb{H}_{A,n}^{\star}$ and $\mathbb{H}_{A,n}^{uc,\star}$.  
Further, estimators for $H_{mix,L}, H_{mix,L}^{uc}, H_{mix,L}^\star$ and $H_{mix,L}^{uc,\star}$ are found by replacing the corresponding arm-specific distribution by their empirical distribution functions and denoted analogously.

\subsection*{Appendix A: Full observations, follow-up time }
In this appendix the single arm trial is considered in follow-up time: In total $n$ patients enter the trial and are followed until occurrence of the the event of interest or censoring at the end of the trial. The aim is to estimate the survival curve at a random moment $\hat t_{n,p}$, the moment $p 100\%$ of the patients has experienced an event, and to derive the asymptotic distribution of the Brelow and the Kaplan-Meier estimator in this point. It is well known that the Kaplan-Meier and the Breslow estimators evaluated in a fixed, non-random, point are asymptotically normal (see e.g.\ Aalen et al.(2008); Therneau and Grambsch, (2000)\nocite{Aalen, Therneau}), but for the random point $\hat t_{p,n}$ the asymptotic distribution is unknown. Note that the setting considered in this appendix is different from the one in Theorem 2; in Theorem 2 (also a one-arm trial) the trial is stopped at the time the interim analysis is performed, at time $\hat t_{p,n}$; the censoring time is random and dependent of the observations and, moreover, possibly not all patients have entered the trial before time $\hat t_{p,n}$. The setting here is simplified on purpose, and the proof in this simplified setting will be used to describe the line of the proof that will also be followed in Appendix B in which Theorem 1 (and Theorem 2) is proved.   

Notation is used as was defined before, but the $A$ in the subscript to indicate the arm of the study, is left out as we consider a one-arm trial here. The observations of the $n$ patients in the study consist of $(T_1\wedge C_1,\Delta_1),\ldots, (T_n\wedge C_n,\Delta_n)$, and are assumed to be iid.  

\bigskip

\noindent
{\bf{\underline{Theorem: single arm, follow-up time}}}\\
Let $\hat S_n$ and $\tilde S_n$ be the Breslow and the Kaplan-Meier estimators of $S$, the survival distribution of the event time, based on all observations until time $L$ (in calendar time): $(T_1\wedge C_1,\Delta_1),\ldots, (T_n\wedge C_n,\Delta_n)$. For $\hat t_{p,n}$ the sample $p$-th quantile of $\mathbb H_n^{uc}$ and $t_p$ the value so that $\hat t_{p,n} \rightarrow t_p$ in probability (as $n\rightarrow \infty$), then $\sqrt{n}(\hat S_n(\hat t_{p,n})-S(t_p)) \leadsto {\cal{N}}(0,\sigma^2)$ and $\sqrt{n}(\tilde S_n(\hat t_{p,n})-S(t_p)) \leadsto {\cal{N}}(0,\sigma^2)$, as $n\rightarrow \infty$, with 
\begin{align}
\sigma^2 &= S(t_p)^2 \int_0^{t_p} \frac{\rd\Lambda(s)}{1-H(s)} \; + \; \frac{f(t_p)^2}{(h^{uc}(t_p))^2} \; p(1-p) \nonumber \\
& \qquad - \; 2S(t_p)\frac{f(t_p)}{h^{uc}(t_p)}\Big((1-p) \Lambda(t_p) +  \int_0^{t_p} \frac{(H^{uc}(s)-H^{uc}(t_p)H(s)) \rd \Lambda(s)}{1-H(s)}\Big).
\label{sigma2E0}
\end{align}

\bigskip

\noindent
{\bf{\underline{Proof: Breslow-estimator, $\hat S_n(\hat t_{p,n})$}}}\\
Write $\sqrt{n}(\hat S_n(\hat t_{p,n})-S(t_p))$ as a sum of three terms:
\begin{align}
\lefteqn{\sqrt{n}(\hat S_n(\hat t_{p,n})-S(t_p))} \nonumber\\ 
&= \; \sqrt{n}(\hat S_n(t_p) - S(t_p)) \; + \; \sqrt{n}(S(\hat t_{p,n}) - S(t_p)) 
\; +\; \sqrt{n}((\hat S_n(\hat t_{p,n})- S(\hat t_{p,n})) - (\hat S_n(t_p) -S(t_p))). 
\label{ThreeTerms}
\end{align}
At the end of this proof, it will be shown that the third term in (\ref{ThreeTerms}) converges in probability to zero. Because of this and application of Slutky's lemma, it follows that it is sufficient to prove that the sum of the first and second term in (\ref{ThreeTerms}), the sum $\sqrt{n}(\hat S_n(t_p) - S(t_p)) \; + \; \sqrt{n}(S(\hat t_{p,n}) - S(t_p))$, is asymptotically normal with mean zero and variance $\sigma^2$. In the following the asymptotic distribution of the two terms are derived separately. Next the sum is considered.

\bigskip

\noindent
{\underline{Asymptotic distribution of the first term: 
$\sqrt{n}(\hat S_n(t_p) - S(t_p))$}}\\
Although the asymptotic distribution of $\sqrt{n}(\hat S_n(t) - S(t))$ for any fixed value of $t$ is known (see e.g.\ Aalen et al.(2008); Therneau and Grambsch, (2000)\nocite{Aalen, Therneau}), the derivation of it is given explicitly, to simplify the proof later on. The key of the proof is the application of the Functional Delta Method (see e.g.\ van der Vaart (1998); van der Vaart and Wellner (1996)\nocite{vdV1998, vdV1996}). 

Because $\mathbb{H}^{uc}_n$ and $\mathbb{H}_n$ are empirical distribution functions of $1\{T_i\wedge C_i\leq t\}$ and $\Delta_i 1\{T_i\wedge C_i \leq t\}$ for $t\geq 0$, the main empirical central limit theorems (Donsker theorem, (see e.g.\ van der Vaart (1998); van der Vaart and Wellner (1996)\nocite{vdV1998, vdV1996}) yield 
\begin{align*}
\sqrt{n}
\begin{pmatrix}
\mathbb{H}^{uc}_n-H^{uc}\\
\mathbb{H}_n-H
\end{pmatrix}
\leadsto 
\begin{pmatrix}
\mathbb G^{uc}\\
\mathbb G
\end{pmatrix}
 \qquad \mbox{in $(D[0,\tau])^2$},
\end{align*}
where $D[0,\tau]$ is the Skorokhod space at $[0,\tau]$ for $\tau>0$ a value so that $(1-F(\tau))(1-G(\tau))>0$ (see e.g.\ van der Vaart (1998) example 20.15; van der Vaart and Wellner, 1996)\nocite{vdV1998,vdV1996} example 3.9.19). The limit $(\mathbb G^{uc},\mathbb G)$ is a tight, zero mean Gaussian process with a covariance structure 
\begin{align}
\label{covariance}
\E \mathbb G^{uc}(s) \mathbb G^{uc}(t) &= H^{uc}(s \wedge t) - H^{uc}(s) H^{uc}(t) \nonumber\\
\E \mathbb G(s) \mathbb G(t) &= H(s \wedge t) - H(s) H(t)\\
\E \mathbb G^{uc}(s) \mathbb G(t) &= H^{uc}(s\wedge t) - H^{uc}(s)H(t). \nonumber
\end{align}

From the expression in (\ref{Ftp}), it can be seen that $S(t_p)$ and $\hat S_n(t_p)$ depend on the pairs $(H^{uc},H)$ and $(\mathbb H_n^{uc},\mathbb H_n)$, respectively, via the sequence of maps, 
\begin{align}
\label{AB}
(A,B) \; \mapsto \; \Big(A,\frac{1}{1-B}\Big) \; \mapsto \;  
\int_0^{t_p} \frac{\rd A}{1-B} \; \mapsto \; \exp\Big(-\int_0^{t_p} \frac{\rd A}{1-B} \Big). 
\end{align}
The composition of these maps is denoted as $z_t$:
\begin{align}
\label{ztp}
z_{t_p}(H^{uc},H) = S(t_p)  \qquad  z_{t_p}(\mathbb{H}_n^{uc},\mathbb{H}_n) = \hat S(t_p).
\end{align}
The composition $z_{t_p}$ is Hadamard-differentiable on $\mathcal{D} := \{(A,B):\int |\rd A| \leq M, 1-B \geq \varepsilon\}$ for given values $M$ and $\varepsilon > 0$, at every pair $(A,B)$ with $1/(1-B)$ of bounded variation. If $t_p$ is in the interval $[0,\tau]$ with $H(\tau)<1$, then the pair of processes $(\mathbb H_n^{uc},\mathbb H_n)$ is contained in $\mathcal{D}$ with probability converging to 1 and with $M\geq 1$ and $\varepsilon$ sufficiently small (van der Vaart (1996)\nocite{vdV1996}, page 384). The derivative map of this composition in (\ref{AB}) is given by
\begin{align*}
(\alpha,\beta) \mapsto -\exp\Big(-\int \frac{\rd A}{1-B}  \Big)\Big( \int \frac{\rd \alpha}{1-B} + \int \frac{\beta \rd A}{(1-B)^2}\Big).
\end{align*}  
By the functional Delta method, it follows that
\begin{align*}
\sqrt{n}(\hat S_n(.)-S(.)) & \leadsto \;
-\exp\Big(-\int \frac{\rd H^{uc}}{1-H} \Big)\Big( \int \frac{\rd \mathbb G^{uc}}{1-H} + \int \frac{\mathbb G \rd H^{uc}}{(1-H)^2}\Big) \qquad \mbox{in $D[0,\tau]$} \nonumber\\
& = \;
-S(.) \Big( \int \frac{\rd \mathbb G^{uc}}{1-H} + \int \frac{\mathbb G \rd H^{uc}}{(1-H)^2}\Big). 
\end{align*}  
Conclude that 
\begin{align}
\sqrt{n}(\hat S_n(t_p)-S(t_p)) & \leadsto \;
-S(t_p) \Big( \int_0^{t_p} \frac{\rd \mathbb G^{uc}}{1-H} + \int_0^{t_p} \frac{\mathbb G \rd H^{uc}}{(1-H)^2}\Big),
\label{SB}
\end{align}
with $\mathbb G^{uc}$ and $\mathbb G$ the Gaussian processes as described before. In the following we will consider this limit distribution. First, define the process
\begin{align*}
\mathbb M_{t_p}(s) \; = \; -S(t_p)\Big(\mathbb G^{uc}(s) + \int_0^s\mathbb G \; \rd \Lambda\Big)
\; = \; -S(t_p) \mathbb M^\star(s), 
\end{align*}
with $\mathbb M^\star(s) = \mathbb G^{uc}(s) + \int_0^s\mathbb G \; \rd \Lambda$.
The process $\mathbb M^\star$ is a zero-mean Gaussian martingale with covariance function
\begin{align*}
\E \mathbb M^\star(s) \mathbb M^\star(t) = \int_0^{s\wedge t} (1-H)\rd\Lambda
\end{align*}
(see van der Vaart (1996)\nocite{vdV1996}, example 3.9.19). 
So, $\mathbb M_{t_p}(.) = -S(t_p) \mathbb M^\star(.)$ is a zero-mean  Gaussian martingale with a covariance function equal to 
\begin{align*}
\E \mathbb M_{t_p}(s) \mathbb M_{t_p}(t) =  S(t_p)^2\int_0^{s\wedge t} (1-H)\rd\Lambda. 
\end{align*}
The limit variable in (\ref{SB}) can be expressed as $\int_{[0,t_p]} (1-H)^{-1}\rd \mathbb M_{t_p}$.
With Martingale theory it can be shown that $\int_{[0,t_p]} (1-H)^{-1}\rd \mathbb M_{t_p}$ is a zero-mean Gaussian process with independent increments and variance 
\begin{align*}
S(t_p)^2\int_0^{t_p} \frac{\rd \Lambda(s)}{(1-F(s))(1-G(s))} \; = \; S(t_p)^2\int_0^{t_p} \frac{\rd \Lambda(s)}{1-H(s)}.
\end{align*}
Combining the results yields 
\begin{align*}
\sqrt{n}(\hat S_n(t_p)-S(t_p)) \; \leadsto \; \mathcal N \Big(0, S(t_p)^2 \int_0^{t_p} \frac{\rd \Lambda(s)}{1-H(s)}\Big).
\end{align*}

\bigskip

\noindent
{\underline{Asymptotic distribution of the second term: 
$\sqrt{n}(S(\hat t_{p,n}) - S(t_p))$}}\\
The asymptotic distribution of sample quantiles is known from the literature. By applying the Delta-method, the asymptotic distribution of $\sqrt{n}(S(\hat t_{p,n}) - S(t_p))$ is found. The proof is given, so that parts can be used later when the asymptotic distribution of the sum of the two terms is determined.

Note that $\hat t_{p,n}$ and $t_p$ are the $p$th sample quantiles of the subdistributions $\mathbb H^{uc}_n$ and $H^{uc}$. For a fixed value $p$, the map that assigns to a cumulative (sub)distribution function its $p$th quantile is Hadamard-differentiable at the domain of distribution functions $\Phi$ that are differentiable at $\Phi^{-1}(p)$ and its derivative $\phi(\Phi^{-1}(p))$ is strictly positive (van der Vaart (1996)\nocite{vdV1996}, example 3.9.21).
Since $\sqrt{n}(\mathbb H^{uc}_n-H^{uc})$ converges in $D(\mathbb R)$ to a Gaussian process, the Functional Delta method yields that 
\begin{align}
\sqrt{n}(\hat t_{p,n}-t_p) \; = \; \sqrt{n}(\mathbb H^{uc,-1}_n(p)-H^{uc,-1}(p)) \; \leadsto  \;
-\frac{\mathbb G^{uc}(t_p)}{h^{uc}(t_p)},
\end{align}
as $n\rightarrow\infty$, a mean zero Gaussian process with variance  
\begin{align*}
\frac{p(1-p)}{(h^{uc}(t_p))^2}.
\end{align*}
Under the assumption that the survival function $S=1-F$ is differentiable in $t_p$, the Delta method (and chain rule) yields that, with $A_{t_p}$ defined as $f(t_p)/h^{uc}(t_p)$,  
\begin{align}
\sqrt{n}(S(\hat t_{p,n})-S(t_p)) \; \leadsto \; \frac{f(t_p)}{h^{uc}(t_p)} \mathbb G^{uc}(t_p) = A_{t_p} \mathbb G^{uc}(t_p),
\end{align}
a normal distribution with mean zero and variance given by $f(t_p)^2 p(1-p)/(h^{uc}(t_p))^2$. 

As a preparation for the next part of the proof, define the composition of these maps as:
\begin{align}
v_{s}(\mu_1,\mu_2) \;=\; S(\mu_1^{-1}(H^{uc}(s))),
\label{vtp}
\end{align}
for $\mu_1$ a non-decreasing function $[0,1]\rightarrow \mathbb R^+$. The inverse $\mu_1^{-1}$ is defined as 
$\mu_1^{-1}(s) = \inf\{u: \mu_1(u)\geq s\}$.
With this definition,
\begin{align*}
v_{t_p}(H^{uc},H) &=\; S(H^{uc,-1}(H^{uc}(t_p)))\;=\; S(t_p),\nonumber\\
v_{t_p}(\mathbb{H}_n^{uc},\mathbb{H}_n) &=\; S(\mathbb H_n^{uc,-1}(H^{uc}(t_p))) \;=\; S(\hat t_{p,n}).
\end{align*}
Note that the second argument in $v_{s}$ is not used, but is included in the definition to simplify notation in the remainder of the proof. 

\bigskip

\noindent
{\underline{Convergence of the third term:
$\sqrt{n}(\hat S_n(\hat t_{p,n})- S(\hat t_{p,n}) - (\hat S_n(t_p)-S(t_p)))$}}\\
It still needs to be proven that the sequence of stochastic variables $\sqrt{n}(\hat S_n(\hat t_{p,n})- S(\hat t_{p,n}) - (\hat S_n(t_p)-S(t_p)))$ converges in probability to zero. Previously, it was shown that $\sqrt{n}(\hat S_n(.)-S(.))$ converges in distribution in $D[0,\tau]$ to a zero-mean Gaussian process with continuous sample paths ($t\rightarrow \Lambda(t)$ is assumed to be continuous). By this continuity and the fact that $\hat t_{p,n}$ converges in probability to $t_p$, it follows that the term in the previous display converges in probability to zero.

\bigskip

\noindent
{\underline{Combining the convergence of the three terms in (\ref{ThreeTerms})}}\\
The third term in (\ref{ThreeTerms}) converges in probability to zero. So, the asymptotic distribution of $\sqrt{n}(\hat S_n(\hat t_{p,n})-S(t_p))$ is determined by the sum of the first and second term in (\ref{ThreeTerms}). In the foregoing the asymptotic distribution of these terms were considered separately, we now have to combine them.  

Define $\phi_t(x,y)=z_t(x,y)+v_t(x,y)$, with $z_t$ and $v_t$ as defined in (\ref{ztp}) and (\ref{vtp}). Then, $\phi_{t_p}(H^{uc},H)=z_{t_p}(H^{uc},H)+v_{t_p}(H^{uc},H)=2S(t_p)$ and $\phi_{t_p}(\mathbb H_n^{uc},\mathbb H_n)=\hat S_n(t_p)+S(\hat t_{p,n})$. That means that
\begin{align*}
\sqrt{n}(\hat S_n(t_p) - S(t_p)) \; + \; \sqrt{n}(S(\hat t_{p,n}) - S(t_p)) \; = \; \sqrt{n}(\phi_{t_p}(\mathbb H_n^{uc},\mathbb H_n) - \phi_{t_p}(H^{uc},H)).
\end{align*}
The map $\phi_{t_p}$ is Hadamard-differentiable since this holds for the maps $z_{t_p}$ and $v_{t_p}$. The derivative $\phi_{t_p}$ equals the sum of the derivatives of $z_{t_p}$ and $v_{t_p}$ and by the Functional Delta Method
\begin{align}
\lefteqn{\sqrt{n}(\hat S_n(t_p) - S(t_p)) \; + \; \sqrt{n}(S(\hat t_{p,n}) - S(t_p))} \nonumber \\
& \leadsto \; 
-S(t_p) \Big(\int_0^{t_p} \frac{\rd \mathbb G^{uc}}{1-H} + \int_0^{t_p} \frac{\mathbb G \rd H^{uc}}{(1-H)^2}\Big) \; + \; \frac{f(t_p)}{h^{uc}(t_p)}\mathbb G^{uc}(t_p)  \qquad \mbox{in $D[0,\tau]$} \nonumber\\
&= \; \int_0^{t_p} \frac{\rd \mathbb M_{t_p}}{1-H} + A_{t_p} \mathbb{G}^{uc}(t_p) 
\; = \; -S(t_p)\int_0^{t_p} \frac{\rd \mathbb M^\star}{1-H} + A_{t_p}\mathbb{G}^{uc}(t_p),
\label{Limdistr}
\end{align}
a zero-mean Gaussian distribution, where $A_{t_p} = f(t_p)/h^{uc}(t_p)$ and $\mathbb M_{t_p}, \mathbb M^\star$ and $\mathbb G^{uc}$ as defined above. This was found by combining the results of the asymptotic theory of the two terms. 
The variance of the limit distribution equals
\begin{align*}
\lefteqn{\var \Big(-S(t_p)\int_0^{t_p} \frac{\rd \mathbb M^\star}{1-H} \; + \; A_{t_p}\mathbb{G}^{uc}(t_p)\Big)}\\ 
&= \var \Big(S(t_p)\int_0^{t_p} \frac{\rd \mathbb M^\star}{1-H}\Big) \; + \; \var (A_{t_p}\mathbb{G}^{uc}(t_p)) \; - \; 2\cov\Big(S(t_p)\int_0^{t_p} \frac{\rd \mathbb M^\star}{1-H},A_{t_p}\mathbb{G}^{uc}(t_p)\Big)\\
&=S(t_p)^2 \int_0^{t_p} \frac{\rd\Lambda}{1-H} \; + \; A_{t_p}^2 p(1-p) \; - \; 2S(t_p)A_{t_p} \cov \Big(\int_0^{t_p} \frac{\rd \mathbb M^\star}{1-H},\mathbb{G}^{uc}(t_p)\Big),
\end{align*}
with 
\begin{align}
\cov \Big(\int_0^{t_p} \frac{\rd \mathbb M^\star}{1-H},\mathbb{G}^{uc}(t_p)\Big) = (1-p) \Lambda(t_p) +  \int_0^{t_p} \frac{(H^{uc}(s)-H^{uc}(t_p)H(s)) \rd \Lambda(s)}{1-H(s)}.
\label{covariance2}
\end{align}
This expression of the covariance follows from
\begin{align*}
\lefteqn{\cov\Big(\int_0^{t_p} \frac{\rd \mathbb M^\star}{1-H}, \mathbb G^{uc}(t_p)\Big) \;=\; \cov\Big(\int_0^{t_p} \frac{\rd \big(\mathbb G^{uc}(s) + \int_0^{s}\mathbb G \; \rd \Lambda \big)}{1-H(s)}, \mathbb G^{uc}(t_p)\Big)}\\
&=\; \cov\Big(\int_0^{t_p} \frac{\rd \mathbb G^{uc}(s)}{1-H(s)}, \mathbb G^{uc}(t_p)\Big) + \cov\Big(\int_0^{t_p} \frac{\mathbb G(s)  \rd \Lambda(s)}{1-H(s)}, \mathbb G^{uc}(t_p)\Big)
\end{align*}
with
\begin{align*}
\cov\Big(\int_0^{t_p} \frac{\mathbb G(s)  \rd\Lambda(s)}{1-H(s)}, \mathbb G^{uc}(t_p)\Big) \;=\; \int_0^{t_p} \frac{\E \mathbb G^{uc}(t_p)\mathbb G(s) \rd \Lambda(s)}{1-H(s)} 
\;=\; \int_0^{t_p} \frac{(H^{uc}(s)-H^{uc}(t_p)H(s))  \rd \Lambda(s)}{1-H(s)}
\end{align*}
and for $0=s_0,\ldots,s_{K_\varepsilon}=t_p$ a grid of width $\varepsilon>0$ at the interval $[0,t_p]$
\begin{align*}
\cov\Big(\int_0^{t_p} \frac{\rd \mathbb G^{uc}(s)}{1-H(s)}, \mathbb G^{uc}(t_p)\Big) 
&=\; \int_0^{t_p} \frac{\E \mathbb G^{uc}(t_p)\rd \mathbb G^{uc}(s)}{1-H(s)}  \\
&=\; \lim_{\varepsilon \downarrow 0}\sum_{i=1}^{K_\varepsilon} \frac{\E \mathbb G^{uc}(t_p)(\mathbb G^{uc}(s_i+\varepsilon)-\mathbb G^{uc}(s_i))}{1-H(s_i)}  \\ 
&=\; \lim_{\varepsilon \downarrow 0}\sum_{i=1}^{K_\varepsilon} \frac{(H^{uc}(s_i+\varepsilon)- H^{uc}(s_i))(1-H^{uc}(t_p))}{1-H(s_i)}   \\
&=\;  (1-H^{uc}(t_p)) \int_{[0,t_p]}\frac{\rd H^{uc}(s)}{1-H(s)}  \;=\;  (1-p) \Lambda(t_p).
\end{align*}

So, the limit distribution is a zero-mean Gaussian  distribution with variance $\sigma^2$ as given in (\ref{sigma2E0}).

\bigskip

\noindent
{\bf{\underline{Proof: Kaplan-Meier estimator, $\tilde S_n(\hat t_{p,n})$}}}\\
The asymptotic distribution of the Kaplan-Meier estimator can be found in a similar way as was found for the Breslow estimator. Again $\sqrt{n}(\tilde S_{n}(\hat t_{p,n})-S(t_p))$ is written as a sum of three terms like in (\ref{ThreeTerms}). For the first term in the decomposition, the asymptotic distribution of the Kaplan-Meier estimator in a fixed point $t_p$ is exactly the same as for the Breslow estimator evaluated in $t_p$. The asymptotic distribution of the second term in (\ref{ThreeTerms}), is also exactly the same as in the first part of the theorem, since it does not depend on the choice of the estimator for $S$. The rest of the proof is exactly as for the Breslow estimator.
\QED\\

\subsection*{Appendix B: Proof of Theorem 1}
In this Appendix we consider the two arm trial again. The trial is stopped at $\hat t_{p,n+m}$ in calendar time, i.e.\ once $p 100\%$ of the patients has experienced an event. This random moment depends on the observations in both arms. Below, the proof of Theorem 1 is given. This proof will follow the same line as was followed in the proof in Appendix A. However, there are some extra complexities. 

\bigskip

\noindent
{\bf{\underline{Theorem 1}}}\\
Let $\hat S_{A,n}$ and $\tilde S_{A,n}$ be the Breslow and Kaplan-Meier estimators for $S_A$, determined based on the observations of the $n$ patients in arm $A$ who entered the study before the time-point $\hat t_{p,n+m}$, the $p$th quantile of the empirical distribution $\mathbb H_{mix,n+m}^{uc,\star}$. Let $\delta>0$ be pre-specified, $\hat t_{p,n+m}^\delta=\hat t_{p,n+m}-\delta$ and its limit $t_p^\delta=t_p-\delta$. The asymptotic distribution of the Breslow estimator and the Kaplan-Meier estimator is
\begin{align*}
\sqrt{n}(\hat S_{A,n}(\hat t_{p,n+m}^\delta)-S_A(t_p^\delta)) &\leadsto {\cal{N}}(0,\sigma_{A,\delta}^2),\\
\sqrt{n}(\tilde S_{A,n}(\hat t_{p,n+m}^\delta)-S_A(t_p^\delta)) &\leadsto {\cal{N}}(0,\sigma_{A,\delta}^2),
\end{align*}
(where $\leadsto$ is the notation for convergence in distribution) as $n,m\rightarrow \infty$ and
\begin{align*}
\sigma_{A,\delta}^2 &= S_A(t_p^\delta)^2 \int_0^{t_p^\delta} \frac{\rd\Lambda_A(s)}{1-H_{A,t_p}(s)} \; + \; \frac{q_A f_A(t_p^\delta)^2}{(h_{mix,t_p}^{uc,\star}(t_p))^2}\; p(1-p) \\
&\qquad -2S_A(t_p^\delta)\frac{q_A\sqrt{q_A}f_A(t_p^\delta)}{h^{uc,\star}_{mix,t_p}(t_p)}\bigg((1 - H_{A,t_p}^{uc,\star}(t_p))\Lambda_A(t_p^\delta)\;+\; \int_0^{t_p^\delta} \frac{(H_{A,t_p}^{uc}(s)-H_{A,t_p}^{uc,\star}(t_p)H_{A,t_p}(s)) \rd\Lambda_A(s)}{1-H_{A,t_p}(s)}\bigg).
\end{align*}

\bigskip

\noindent
{\bf{\underline{Proof: Breslow estimator, $\hat S_{A,n}(\hat t_{p,n+m})$}}}\\
First, the sequence $\sqrt{n}(\hat S_{A,n}(\hat t_{p,n+m}^{\delta})-S_A(t_p^{\delta}))$ is decomposed as a sum of three terms of which the first term deals with the estimation of $S_A$ at a fixed point $t_p^\delta$, the second term with the estimation of the time $S_A(t_p^\delta)$ by $S_A(\hat t_{p,n+m}^\delta)$ and a third term that converges to zero in probability:
\begin{align}
\label{ThreeTerms2}
\sqrt{n}(\hat S_{A,n}(\hat t_{p,n+m}^\delta)-S_A(t_p^\delta))    
& = \; \sqrt{n}(\hat S_{A,n}(t_p^\delta) - S_A(t_p^\delta)) \; + \; \sqrt{n}(S_A(\hat t_{p,n+m}^\delta) - S_A(t_p^\delta))    \nonumber\\
& +\; \sqrt{n}((\hat S_{A,n}(\hat t_{p,n+m}^\delta)- S_A(\hat t_{p,n+m}^\delta)) - (\hat S_{A,n}(t_p^\delta) -S_A(t_p^\delta))). 
\end{align}
The main difficulty when deriving the asymptotic distribution of the term $\sqrt{n}(\hat S_{A,n}(t_{p}^{\delta})-S_A(t_p^{\delta}))$, is the fact that the moment the interim analysis is performed, $\hat t_{p,n+m}$, is stochastic and is a function of all observations. 
To deal with this, take $\varepsilon>0$ so that $0<\varepsilon<\delta$. The value $t_p-\varepsilon$ is fixed and non-stochastic and is (for the moment) taken as the moment of the new ``end-of-study''. The distribution for $C_A$ now equals $G_{t_p-\varepsilon}$. It might be that $G_{Acc}(t_p-\varepsilon)<1$, meaning that possibly not all patients have entered the trial at the time of the interim analysis, and making the sample size random. This problem can be circumvented by redefining $G_{Acc}$ so that it has a point mass at $t_p-\varepsilon$ of size $1-G_{Acc}(t_p-\varepsilon)$ and assuming that $G_{t_p-\varepsilon}$ has a point mass at 0, also of size $1-G_{Acc}(t_p-\varepsilon)$. This means that possibly a fraction of the patients enter the study at the moment the trial is stopped and have a follow-up time of 0. By this redefinition, the sample size is fixed at $n$ again, but possibly a fraction of the patients do not contribute any information on the survival curve (have follow-up time 0).
With this, the asymptotic distribution of $\hat S_{A,n}(\hat t_p^\delta)$ can be derived along the same lines as in the proof in Appendix A. Conclude that the asymptotic distribution of the first term in (\ref{ThreeTerms2}), in case of censoring at $t_p-\varepsilon$ is asymptotically normal with mean zero and variance $S_A(t_p^\delta)^2 \int_0^{t_p^\delta} (1-H_{A,t_p-\varepsilon})^{-1}\rd\Lambda_A$. The same reasoning can be done for the new end-point $t_p+\varepsilon$. This yields an asymptotic normal distribution with mean zero and variance $S_A(t_p^\delta)^2 \int_0^{t_p^\delta} (1-H_{A,t_p+\varepsilon})^{-1}\rd\Lambda_A$. Next, remind that $1-H_{A,t_p+\varepsilon}(s) = S_A(s)G_{Acc}(t_p+\varepsilon-s)$ is increasing in $\varepsilon$ and the asymptotic variance is decreasing with $\varepsilon$. This is as expected, because the longer the follow-up, the more information and the smaller the asymptotic variance. For $\hat S_{A,n}^{t_p-\varepsilon}$ and $\hat S_{A,n}^{t_p+\varepsilon}$ the Breslow estimators based on data that are censored at $t_p-\varepsilon$ and $t_p+\varepsilon$, respectively, and for $t>0$ 
\begin{align*} 
\lefteqn{\lim_{n\rightarrow \infty}\rP(\sqrt{n}(\hat S_{A,n}^{t_p-\varepsilon}(t_p^\delta)-S_A(t_p^\delta))\leq t) }\\
&=\; \lim_{n,m\rightarrow \infty}\rP(\sqrt{n}(\hat S_{A,n}^{t_p-\varepsilon}(t_p^\delta)-S_A(t_p^\delta))\leq t, t_p-\varepsilon \leq \hat t_{p,n+m} \leq t_p+\varepsilon) \\
&\leq\; \liminf_{n,m\rightarrow \infty}\rP(\sqrt{n}(\hat S_{A,n}^{\hat t_{p,n+m}}(t_p^\delta)-S_A(t_p^\delta))\leq t, t_p-\varepsilon \leq \hat t_{p,n+m} \leq t_p+\varepsilon) \\
&\leq\; \limsup_{n,m\rightarrow \infty}\rP(\sqrt{n}(\hat S_{A,n}^{\hat t_{p,n+m}}(t_p^\delta)-S_A(t_p^\delta))\leq t, t_p-\varepsilon \leq \hat t_{p,n+m} \leq t_p+\varepsilon) \\
&\leq\; \lim_{n,m\rightarrow \infty}\rP(\sqrt{n}(\hat S_{A,n}^{t_p+\varepsilon}(t_p^\delta)-S_A(t_p^\delta))\leq t, t_p-\varepsilon \leq \hat t_{p,n+m} \leq t_p+\varepsilon) \\
&=\; \lim_{n\rightarrow \infty}\rP(\sqrt{n}(\hat S_{A,n}^{t_p+\varepsilon}(t_p^\delta)-S_A(t_p^\delta))\leq t),
\end{align*}
because of the relationship between $\varepsilon$ and the asymptotic variance. For $t<0$, the same reasoning holds with the inequality signs reversed. Now, let $\varepsilon$ decrease to zero to conclude that 
\begin{align*}
\sqrt{n}\bigg(\hat S_{A,n}^{\hat t_{p,n+m}}(t_p^\delta)-S_A(t_p^\delta)\bigg) \leadsto {\cal{N}}\bigg(0,S_A(t_p^\delta)^2 \int_0^{t_p^\delta} (1-H_{A,t_p})^{-1}\rd\Lambda_A\bigg).    
\end{align*}

For the derivation of the asymptotic distribution of the second term in (\ref{ThreeTerms2}), the line of the proof in Appendix A can be followed. The moment $\hat t_{p,n+m}$ is the quantile of the empirical distribution function $\mathbb H^{uc,\star}_{mix}$, for the combined sample of sample size $n+m$. Write, for $q_A=n/(n+m)$
\begin{align*}
\sqrt{n}(S_A(\hat t_{p,n+m}^\delta)-S_A(t_p^\delta)) 
&=\; -\sqrt{q_A} \sqrt{n+m}(\hat t_{p,n+m}^\delta-t_p^\delta)f_A(t_p^\delta) + o_P(1) \\
&=\; -\sqrt{q_A} \sqrt{n+m}(\hat t_{p,n+m}-t_p)f_A(t_p^\delta) + o_P(1)
\end{align*}
where the first equality is the a direct consequence of a first Taylor expansion (or the Delta method).
The last term is asymptotically normal with mean zero and variance equal to the second term in $\sigma^2_{A,\delta}$, as was seen in the proof in Appendix A. So: 
\begin{align*}
\sqrt{n}(S_A(\hat t_{p,n+m}^\delta)-S_A(t_p^\delta)) 
\; \leadsto \; \sqrt{q_A} \frac{f_A(t_p^\delta)}{h^{uc,\star}_{mix,t_p}(t_p)} {\mathbb G}^{uc,\star}_{mix}(t_p),
\end{align*}
with ${\mathbb G}^{uc,\star}_{mix}(t_p)$ a mean zero Gaussian process with variance $p(1-p)$ (as before).

The third term in (\ref{ThreeTerms2}) converges in probability to zero, as was seen in Appendix A.

\medskip

Now we need to combine the asymptotic distribution of the three terms again. This needs some extra arguments as the derivation of the second term was based the mixture distribution. We have seen that, like in the proof in Appendix A, 
\begin{align*}
\sqrt{n}(\hat S_{A,n}(t_p^\delta) - S_A(t_p^\delta)) \; + \; \sqrt{n}(S_A(\hat t_{p,n+m}^\delta) - S_A(t_p^\delta))
\; \leadsto \; 
 -S_A(t_p^\delta)\int_0^{t_p^\delta} \frac{\rd \mathbb M^\star}{1-H_{A,t_p}} \; + \; \sqrt{q_A} \frac{f_A(t_p^\delta)}{h^{uc,\star}_{mix,t_p}(t_p)}\mathbb{G}^{uc,\star}_{mix}(t_p),
\end{align*}
a zero-mean Gaussian distribution (see the proof in Appendix A), with $\mathbb M_{t_p}^\star$ and $\mathbb G^{uc,\star}_{mix}$ as defined before.
The variance of the distribution is given by
\begin{align*}
\lefteqn{\var \Big(-S_A(t_p^\delta)\int_0^{t_p^\delta} \frac{\rd \mathbb M^\star}{1-H_{A,t_p}} \; + \; \sqrt{q_A}\frac{f_A(t_p^\delta)}{h^{uc,\star}_{mix,t_p}(t_p)}\mathbb{G}^{uc,\star}_{mix}(t_p)\Big)}\\ 
&= \var \Big(S_A(t_p^\delta)\int_0^{t_p^\delta} \frac{\rd \mathbb M^\star}{1-H_{A,t_p}}\Big) \; + \; \var \Big(\frac{ \sqrt{q_A}f_A(t_p^\delta)}{h^{uc,\star}_{mix,t_p}(t_p)}\mathbb{G}_{mix}^{uc,\star}(t_p)\Big) \\
&\qquad - \; 2\cov\Big(S_A(t_p^\delta)\int_0^{t_p^\delta} \frac{\rd \mathbb M^\star}{1-H_{A,t_p}},\frac{\sqrt{q_A} f_A(t_p^\delta)}{h^{uc,\star}_{mix,t_p}(t_p)}\mathbb{G}^{uc,\star}_{mix}(t_p)\Big)\\
&= S_A(t_p^\delta)^2 \int_0^{t_p^\delta} \frac{\rd\Lambda_A}{1-H_{A,t_p}} \; + \; q_A\Big(\frac{f_A(t_p^\delta)}{h^{uc,\star}_{mix,t_p}(t_p)}\Big)^2 p(1-p) \\ 
& \qquad - \; 2\sqrt{q_A} \frac{ S_A(t_p^\delta) f_A(t_p^\delta)}{h^{uc,\star}_{mix,t_p}(t_p)}\; \cov \Big(\int_0^{t_p^\delta} \frac{\rd \mathbb M^\star}{1-H_{A,t_p}},\mathbb{G}^{uc,\star}_{mix}(t_p)\Big),
\end{align*}
with
\begin{align*}
\lefteqn{\cov\Big(\int_0^{t_p^\delta} \frac{\rd \mathbb M^\star}{1-H_{A,t_p}}, \mathbb G^{uc,\star}_{mix}(t_p)\Big) \;=\; \cov\Big(\int_0^{t_p^\delta} \frac{\rd \big(\mathbb G_A^{uc}(s) + \int_0^{s}\mathbb G_A \; \rd \Lambda \big)}{1-H_{A,t_p}(s)}, \mathbb G^{uc,\star}_{mix}(t_p)\Big)}\\
&=\; \cov\Big(\int_0^{t_p^\delta} \frac{\rd \mathbb G^{uc}_A(s)}{1-H_{A,t_p}(s)}, \mathbb G_{mix}^{uc,\star}(t_p)\Big) + \cov\Big(\int_0^{t_p^\delta} \frac{\mathbb G_A(s)  \rd \Lambda(s)}{1-H_{A,t_p}(s)}, \mathbb G^{uc,\star}_{mix}(t_p)\Big)\\
&=\; q_A\cov\Big(\int_0^{t_p^\delta} \frac{\rd \mathbb G^{uc}_A(s)}{1-H_{A,t_p}(s)}, \mathbb G_{A}^{uc,\star}(t_p)\Big) + q_A\cov\Big(\int_0^{t_p^\delta} \frac{\mathbb G_A(s)  \rd \Lambda(s)}{1-H_{A,t_p}(s)}, \mathbb G^{uc,\star}_{A}(t_p)\Big),
\end{align*}
where the last equality follows from the independence between the observations in the two arms. The covariances in the last line in the previous display equal
\begin{align*}
\cov\Big(\int_0^{t_p^\delta} \frac{\mathbb G_A(s)  \rd\Lambda_A(s)}{1-H_{A,t_p}(s)}, \mathbb G^{uc,\star}_{A}(t_p)\Big) &=\; \int_0^{t_p^\delta} \frac{\E \mathbb G^{uc,\star}_{A}(t_p)\mathbb G_A(s) \rd \Lambda_A(s)}{1-H_{A,t_p}(s)}\\ 
&=\; \int_0^{t_p^\delta} \frac{(H_{A,t_p}^{uc}(s)-H_{A,t_p}^{uc,\star}(t_p)H_{A,t_p}(s))  \rd \Lambda_A(s)}{1-H_{A,t_p}(s)}
\end{align*}
and in a similar way as in Appendix A, it is found that
\begin{align*}
\cov\Big(\int_0^{t_p^\delta} \frac{\rd \mathbb G_A^{uc}(s)}{1-H_{A,t_p}(s)}, \mathbb G^{uc,\star}_{A}(t_p)\Big) \;=\; (1-H_{A,t_p}^{uc,\star}(t_p))\Lambda_A(t_p^\delta).
\end{align*}
 

\bigskip

\noindent
{\bf{\underline{Proof: Kaplan-Meier estimator, $\tilde S_{A,n}(\hat t_{p,n+m})$}}}\\
This can be done in the same line as was done for the Breslow-estimator.
\QED

\end{document}